\documentclass[amssymb,amsmath,aps,showpacs,floatfix,nofootinbib,showpacs,12pt]{revtex4}
\usepackage{graphicx,epsfig}
\def\beq{\begin{equation}}
\def\eeq{\end{equation}}
\def\be{\begin{equation}}
\def\ee{\end{equation}}
\def\bea{\begin{eqnarray}}
\def\eea{\end{eqnarray}}
\def\nnb{\nonumber}

\newcommand{\gsim}{\lower.7ex\hbox{$\;\stackrel{\textstyle>}{\sim}\;$}}
\newcommand{\lsim}{\lower.7ex\hbox{$\;\stackrel{\textstyle<}{\sim}\;$}}

\begin{document}


 \title{ Electron events from the scattering with solar neutrinos 
 in the search of keV scale sterile neutrino dark matter}
 \author{ Wei Liao, Xiao-Hong Wu and Hang Zhou}
 \affiliation{
  Institute of Modern Physics, School of Sciences \\
 East China University of Science and Technology, \\
 130 Meilong Road, Shanghai 200237, P.R. China %
}


\begin{abstract}
 In a previous work we showed that keV scale sterile neutrino dark 
 matter $\nu_s$ is possible to be detected in $\beta$ decay experiment
 using radioactive sources such as $^3$T or $^{106}$Ru. The signals
 of this dark matter candidate are mono-energetic electrons produced
 in neutrino capture process $\nu_s + N' \to N +e^-$. These electrons
 have energy greater than the maximum energy of the electrons produced
 in the associated decay process $N' \to N +e^- + {\bar \nu}_e$. Hence, signal 
 electron events are well beyond the end point of the $\beta$ decay 
 spectrum and are not polluted by the $\beta$ decay process. Another 
 possible background, which is a potential threat to the detection of
 $\nu_s$ dark matter, is the electron event produced by the scattering 
 of solar neutrinos with electrons in target matter. In this article 
 we study in detail this possible background and discuss its implication 
 to the detection of keV scale sterile neutrino dark matter. In particular, 
 bound state features of electrons in Ru atom are considered with care
 in the scattering process when the kinetic energy of the final electron 
 is the same order of magnitude of the binding energy.
\end{abstract}
 \maketitle

 {\bf Introduction}

 Among many dark matter(DM) candidates, keV scale sterile neutrino
 warm DM is a very interesting possibility. It has several virtues. Among
 them include 1) it's capable to smooth the structure of the universe 
 at small scale and reduce the over-abundance of small scale structures
 observed in the simulation of cold DM scenarios~\cite{structure};
 2) it provides a fermionic dark matter candidate with an appropriate 
 mass scale which naturally avoids the cusp core in
 the halo density profile~\cite{Cusp}; 3) it naturally gives a lifetime 
 longer than the age of the universe and does not need to include by hand a 
 discrete or global symmetry to guarantee the stability or long lifetime 
 of the DM~\cite{Liao,Liao2}; 4) it's easy to get such a DM candidate
 in well motivated models such as the seesaw models~\cite{ABS, review2}
 which are the most popular models for explaining the tiny masses
 of active neutrinos.

 Many aspects of this warm DM candidate, e.g. the production
 mechanism in the early universe~\cite{DW,ABS,others,blrv,ST,PK}, the 
 astrophysical and cosmological constraints, possible
 models and symmetries, etc., have been analyzed and considered
 ~\cite{bhl,bnrst,bnr,birs,blrv2,bri,sh,LMN,GT,CDS,RT}
 model-independently or in special models. Among them, attention
 has been paid to the detection of this keV scale DM.
 It was realized that indirect detection of this DM background
 in the universe to a good sensitivity can be achieved in principle
 using satellite observation of mono-energetic X-rays produced in
 two-body decay of the DM: $\nu_s \to \nu +\gamma$. However this observation
 scheme requires large statistics which is not available
 in present scale satellite observation program~\cite{indirect}.
 Direct detection of this DM candidate in laboratory has
 also been investigated~\cite{Liao,Liao2,LX,LX1,SV,BS,ak}. 
 Because of its small mass and weak interaction, some authors found it not 
 possible to detect this
 DM candidate in laboratory~\cite{LX1,BS,ak,SV}.

 In a previous work we proposed that keV scale $\nu_s$ DM
 can be detected using $\beta$ decay nuclei such as $^3$T
 and $^{106}$Ru~\cite{Liao}. It was found that with a small mixing with
 electron neutrino $\nu_e$, $\nu_s$ can be captured by
 $^3$T and $^{106}$Ru in process $\nu_s + N' \to N +e^-$.
 The signals of $\nu_s$ DM are mono-energetic electrons 
 with energy well beyond the end point of the $\beta$ decay spectrum.
 It was shown in \cite{Liao} that the signal electrons
 produced in the $\nu_s$ capture process have kinetic energy
 $Q_\beta+m_{\nu_s}$ where $m_{\nu_s}$ is the mass of $\nu_s$.
 $Q_\beta$ is the decay energy of the decay process
 $N'\to N+e^- +{\bar \nu}_e$. $Q_\beta$ equals to the maximal
 kinetic energy of electrons produced in the decay process,
 i.e. the kinetic energy at the end point of the decay spectrum.
 For $^{106}$Ru, $Q_\beta =39.4$ keV. For $^3$T, $Q_\beta =18.59$ keV. 
 So for $m_{\nu_s}=2-5$ keV, the signal electrons have
 kinetic energy around $20$ keV for $^3$T and around $40$ keV for
 $^{106}$Ru respectively.
 We found that with reasonable amount of $^3$T and $^{106}$Ru target 
 we can get a few to tens signal electrons per year. 
 Hence, we concluded that detection of keV scale sterile neutrino DM is 
 possible using this detection scheme. This conclusion is
 of general significance and the detection scheme using
 $\beta$ decay nuclei can be applied to $\nu_s$ DM in 
 different models. More details and general features of 
 this detection scheme have been further analyzed in~\cite{LX}.
 
 Two kinds of background electron events have also been
 discussed in ~\cite{Liao}. One type of background events
 are generated when radioactive nuclei $^3$T or $^{106}$Ru
 capture solar electron neutrinos of energy around keV 
 and produce final electrons in an energy range close to that
 of $\nu_s$ capture process. These electrons can mimick the
 signal electron of $\nu_s$ DM. Fortunately, this type of
 background is sufficiently small because solar neutrino
 flux at keV energy range is pretty small. Another possible 
 background events are electrons kicked out by neutrinos
 in the scattering of 
 solar neutrinos with electrons in target matter. This type
 of background should be discussed in detail, which however was not
 done in detail in \cite{Liao}, for the following reasons: 1) solar
 neutrinos with higher energy can also participate in the scattering
 process and the total solar neutrino flux contributing to the 
 background events is significantly higher than that for the
 first type background; 2) since signal electrons have
 energy of tens keV which is the same order
 of magnitude of the binding energy of the electrons in inner shell
 of $^{106}$Ru atoms, the bound state feature of initial 
 electrons in the scattering process should be taken into
 account at least for the scattering with electrons in
 inner shell of $^{106}$Ru atoms. 

 In this article we will discuss in detail the scattering of
 neutrinos with bound state electrons. Attention will be
 paid to the scattering of neutrinos with electrons in inner shell
 of Ru atom and the events of final electrons with energy of tens keV.
 We will analyze the detailed energy distribution of the
 scattering of neutrinos with bound state electrons.
 We will discuss the dependence of the scattering process on the neutrino
 energy. We will analyze electron events caused by the scattering with
 solar neutrinos and the limitations for the detection
 of keV scale sterile neutrino DM.
 In the following we will first review the bound state feature
 of electrons in $^{106}$Ru and discuss some general features
 of the scattering of neutrinos with bound electrons.
 Then we will come to detailed discussions of the scattering of
 neutrinos with electrons in $^{106}$Ru. Finally, we will discuss the
 events of the scattering of solar neutrinos with bound state electrons
 and discuss the implication for DM detection.

 {\bf Electrons in $^{106}$Ru atoms and its interaction with
 neutrinos}

 In this section we briefly review features of bound electrons
 in $^{106}$Ru atom and discuss some general features of the
 scattering of neutrinos with bound electrons. For reasons
 to be explained below, we will not discuss in detail the 
 scattering with bound electrons in $^3$T atom.

 As noted above, the signal electrons produced in $\nu_s$ DM capture process,
 $\nu_s +N' \to N +e^-$, have kinetic energy $Q_\beta +m_{\nu_s}$.
 For a keV scale $\nu_s$ DM with a mass $2-5$ keV, 
 the signal electrons have kinetic energy
 around $20$ keV for $^3$T and $40$ keV for $^{106}$Ru separately. 
 
 The atomic number of the Ruthenium element is 44. 
 In Table. \ref{Tab:Ru-atom} we list the electronic levels of the 
 ground state configuration of neutral Ruthenium atom~\cite{CRC}
 and the corresponding binding energies obtained from X-ray
 data~\cite{CRC,X-ray-evaluation}.
 One can see that electrons in $K$ and $L$ shells have
 binding energies greater than keV. When considering final
 electrons with energy round $40$ keV, one would expect that 
 bound state features may give some effects to the scattering
 of neutrinos with electrons at these energy levels.
 We would also expect that the effect of the bound state features
 in the scattering with electrons in L shell should be
 weaker than that in the scattering with electrons
 in K shell.

 \begin{table}
\begin{center}
\begin{tabular}{|c|c|c|c|c|c|c|c|c|c|c|c|c|c|c|c|}
\hline
Electronic level & $K$ & $L_I$ & $L_{II}$ & $L_{III}$
& $M_I$ & $M_{II}$ & $M_{III}$ & $M_{IV}$
& $M_{V}$ & $N_I$ & $N_{II}$ & $N_{III}$ & $N_{IV}$ & $N_{V}$ & $O_I$
\\
\hline
State & $1s$ & $2s$ & $2p_{\frac{1}{2}}$ & $2p_{\frac{3}{2}}$ & $3s$ &
 $3p_{\frac{1}{2}}$ & $3p_{3 \over 2}$ & $3d_{3 \over 2}$ & $3d_{5 \over 2}$ &
 $4s$ & $4p_{1\over 2}$ & $4p_{3 \over 2}$ & $4d_{3 \over 2}$ &
 $4d_{5 \over 2}$ & $5s$
\\
\hline
No. of electrons & 2 & 2 & 2 & 4 & 2 & 2 & 4 & 4 & 6 &
2& 2 & 4 & 4 & 3 & 1
\\
\hline
Binding E(keV) & $22.1$ & $3.22$ & $2.97$ & $2.84$ & $0.59$ & $0.48$ &
$0.46$ & $0.28$ & $0.28$ & $0.075$ & $0.043$ & $0.043$ & $0.002$ & $0.002$ &
\\
\hline
  \end{tabular}
 \end{center}
\caption{\it Ground state configuration of neutral $Ru$ atom
and the binding energy from X-ray experiment~\cite{X-ray-evaluation,CRC}.}
\label{Tab:Ru-atom}
\end{table}

 For the scattering with electrons in M, N and O shells in Ru atom, 
 we expect that
 bound state features of these electrons
 should not give large correction to the 
 neutrino-electron scattering of interests to us.
 As noted above, we would be interested in the events
 of electrons with energy around $40$ keV. As can be seen
 in Table \ref{Tab:Ru-atom}, this energy is at least about two orders
 of magnitude larger than the binding energy of electrons
 in M, N and O shells. Detailed studies, to be given in the following,
 shows that even effects of the bound state feature of the
 electrons in L shell are not large in the energy
 range $\gsim 40$ keV. These studies support what we expect
 for the scattering with electrons in M, N and O shells.
 
 For $^3$T, the binding energy is the famous $13.6$ eV.
 It is three orders of magnitude smaller than
 the kinetic energy of interests to us, say $\sim 20$ keV.
 Similar to the discussion above for the electrons in M, N and O
 shells of Ru atom, it's natural to expect that
 bound state features of initial electron in $^3$T should not
 give significant effect to the neutrino-electron scattering process when
 discussing events with final electrons of an energy around $20$ keV.

 To understand in detail the scattering of neutrino with bound
 electron in Ru atom we need the wavefunction of the bound electron.
 Neglecting relativistic corrections, interactions of 
 non-relativistic electrons in an atom include interaction with the
 nucleus through a central force and the interaction
 with other electrons. 
 The Hamiltonian for such a system can be written as
 \bea
 {\cal H}=\sum_i \frac{{\hat p_i}^2}{2 m_e}-\sum_i\frac{Z e^2}{r_i}
 +\sum_{i<j}\frac{e^2}{r_{ij}},
 \eea
 where ${\hat p}_i$ is the momentum operator of the $i$th electron,
 $r_i$ the radius of the $i$th electron, $r_{ij}$ the 
 distance between $i$th electron and $j$th electron,
 $e$ the charge of electron, $Z$ the atomic number.
 It's very difficult to solve
 states of electrons including all details of these interactions.
 Fortunately, one can take the approximation that 
 the forces acted by all other electrons on a single electron 
 can be approximated to be a mean central force and the
 state of a single electron can be solved using an effective
 Hamiltonian
 \bea
 {\cal H}_i= \frac{{\hat p_i}^2}{2 m_e}-\frac{Z e^2}{r_i}
 +V_i(r_i),
 \eea
 where $V_i$ arises from the interaction of $i$th electron
 with all other electrons. A further approximation
 one can use is that the effect of an electron in
 an outer shell on an electron in an inner shell can be
 taken small, since the distance between these electrons
 can be considered large compared to distances between
 electrons in inner shell.

 For electron in K shell of Ru atom, the above approximation
 works well. According to the approximation described above,
 an electron in K shell is similar to an electron in the ground
 state of the hydrogen atom except that the atomic number is replaced
 by $Z=44$. Using this approximation, the binding energy of an 
 electron in K shell is estimated to be $\varepsilon=m_e Z^2 \alpha^2/2$ 
 where $\alpha$ is the fine structure constant. Using this approximation
 for an electron in K shell of Ru atom, we find that 
 $\varepsilon\approx 26.3$ keV
 which is consistent with the binding energy from X-ray data
 shown in Table \ref{Tab:Ru-atom}. This convinces us that the
 wavefunction of an electron in K shell of Ru atom can be
 approximated as the one similar to the wavefunction of 1s state
 in hydrogen atom. In momentum space this wavefunction
 can be written as
 \bea
 \varphi_{1s}(p)= \frac{2\sqrt{2}}{\pi}\frac{a_K^{3/2}}{(p^2a_K^2+1)^2}.
 \label{wavefunction-1s}
 \eea
 It satisfies
 \bea
 \int ~d^3 p~ |\varphi_{1s}(p)|^2 =1. \label{wavefunction-1sa}
 \eea
 Eq. (\ref{wavefunction-1s}) is normalized to give a
 kinetic energy $1/(2 m_e a_K^2)$ where 
 $a_K$ is the effective radius of the electron in
 the state of K shell. The binding energy equals to the
 kinetic energy $\varepsilon_K=1/(2 m_e a_K^2)$, as
 a consequence of the Virial theorem. So $a_K$ can be determined
 using binding energy shown in Table \ref{Tab:Ru-atom}.
 If using only the central force from the
 interaction with the nucleus one has $a_K=1/(m_e Z \alpha)$
 and one can recover the previous estimate:$\varepsilon_K=m_e Z^2 \alpha^2/2$.

 For electrons in L shell of Ru atom, the above approximation
 seems also working well. The evidence is the quasi-degeneracy
 of binding energies of $L_I$, $L_{II}$ and $L_{III}$ states as 
 shown in Table \ref{Tab:Ru-atom}.
 One can see that the binding energies of these energy levels
 are almost the same. This suggests that the dynamical $SO(4)$
 symmetry works well for electrons in these states and the mean
 potentials acted on these electrons should be close to a $1/r$ 
 law. So the wavefunctions of electrons in L level can be
 approximated as that similar to the $2s$ and $2p$ wavefunctions
 in the hydrogen atom. In momentum space we write these 
 wavefunctions as
 \bea
 \varphi_{2s}(p) &&=\frac{a_L^{3/2}}{\pi}\frac{p^2 a_L^2-1/4}{(p^2a_L^2+1/4)^3},
 \label{wavefunction-2s} \\
 \varphi_{2p0}(p) &&=-\frac{i a_L^{3/2}}{\pi}\frac{p_z a_L}{(p^2 a_L^2+1/4)^3}, 
 \label{wavefunction-2p0}\\
 \varphi_{2p\pm 1}(p)&&=\frac{a_L^{3/2}}{\sqrt{2}\pi}\frac{(p_x \pm i p_y)a_L}
 {(p^2a_L^2+1/4)^3},
 \label{wavefunction-2p1}
 \eea
 where $2pi$($i=0,\pm1$) refers to states with different
 projections of angular momentum onto z direction.
 These wavefunctions satisfy
 \bea
 \int ~d^3p ~ |\varphi_l(p)|^2=1, ~~\textrm{$l=2s, 2p0, 2p+1,2p-1$}.
 \eea
 They are normalized to give a kinetic energy $1/(2^3 m_e a_L^2)$.
 The binding 
 energy equals to the kinetic energy $\varepsilon_L=1/(2^3 m_e a_L^2)$,
 as a consequence of the Virial theorem. So $a_L$ can be determined
 using binding energy shown in Table \ref{Tab:Ru-atom}.
 For simplicity we use a universal $\varepsilon_L$ for all 
 $2s$ and $2p$ states in definitions given
 in Eqs. (\ref{wavefunction-2s}), (\ref{wavefunction-2p0})
 and (\ref{wavefunction-2p1}).

 For electrons in M, N and O shells, the situation is a bit
 complicated. As can be seen in Table \ref{Tab:Ru-atom},
 there are no quasi-degeneracies of states in M, N and O shells.
 Fortunately, these states have much smaller binding energies
 than the states in K and L shells. In particular, their binding
 energies are about two orders of magnitude smaller than the energy
 range of the kinetic energy of scattered electrons, i.e. $\gsim 40$ keV,
 which is of interests to us. We would expect that the bound state
 features of electrons in these states should not alter the
 scattering process significantly and we should be able to
 approximate these electrons as at rest with the energy equals to
 the rest mass.

{\bf Scattering of neutrino with bound electrons in Ru atom}

 For the scattering of a neutrino with free electron at rest,
 the cross section is well known~\cite{RPP}:
 \bea
 \frac{d\sigma_0}{dE_k}=\frac{G_F^2m_e}{2\pi} \bigg[ (g_V+g_A)^2
 +(g_V-g_A)^2(1-\frac{E_k}{E_\nu})^2
 -(g_V^2-g_A^2)\frac{E_k m_e}{E_\nu^2} \bigg],
 \label{X-sec0}
 \eea
where $E_\nu$ is the energy of initial neutrino,
$E_k$ the kinetic energy of scattered electron,
$g_{V,A}$ are the vector and axial-vector coupling constants~\cite{RPP}.
$g_V=-\frac{1}{2}+2\sin^2\theta_W$ and $g_A=-\frac{1}{2}$
for muon and tau neutrino. For electron neutrino
$g_V=\frac{1}{2}+2\sin^2\theta_W$, $g_A=\frac{1}{2}$.
$\theta_W$ is the weak mixing angle with $\sin^2\theta_W=0.231$.
The kinetic energy of the scattered electron lies in the range 
\bea
E_k \subset [0, E_\nu/(1+m_e/(2E_\nu))]. \label{range0}
\eea
 One can see in Eq. (\ref{X-sec0}) that the differential
cross section varies slowly with respect to $E_k$.

 For the scattering of a neutrino with bound electron
 in Ru atom, the neutrino
 directly transfers energy and momentum to the electron. The neutrino
 neither affects the other parts of the atom nor is affected by the
 other parts of the atom. This is the exactly the case that
  the impulse approximation works~\cite{PWIA}. Furthermore,
 we can approximate the wavefunction of the scattered electron 
 as the plane wave. This is because we would concentrate on
 the energy range, $\gsim 40$ keV, for the scattered electron.
 For kinetic energy larger than the binding energy one can take the
 plane wave approximation for the wavefunction of 
 the scattered electrons~\cite{PWIA}.

 The scattering of a neutrino with bound electrons
 should in principle be treated in a relativistic framework.
 This requirement would give rise to complication if
 taking into account relativistic wavefunction of electron
 in bound state. Fortunately, we can simplify the discussion
 by noticing that the electrons in Ru atoms can be
 considered non-relativistic, as can be seen in Table \ref{Tab:Ru-atom}.
 Furthermore, the spin-orbit coupling is zero for electrons
 in K shell which have zero angular momentum and the energy shift 
 due to the spin-orbit coupling to electrons in L shell should give
 negligible effect to our result for the events with scattered electrons
 of $E_k \gsim 40$ keV. So we can approximately take spin and
 angular momentum as independent variables, as
 in the case of non-relativistic quantum mechanics. In this approximation
 the total wavefunction of the bound electron 
 can be approximated as a product of a spinor and a wavefunction in x or p
 coordinates:
 \bea
 \Psi({\vec x})=\psi \phi({\vec x}),~~\Phi({\vec p})= \psi \varphi({\vec p}),
 \label{chiral}
 \eea
 where $\psi$ is a spinor and it equals to $(1,0,0,0)^T$ or $(0,1,0,0)^T$
 in standard Dirac representation of spinor.

 In this approximation, one can find that
 the cross section of the scattering of a neutrino with an electron 
 in a particular state $l$ can be written as
 \bea
 \frac{d \sigma_l}{d E_k}= \int~d^3p_B~
 |\varphi_l({\vec p}_B)|^2 \frac{d\sigma_{{\vec p}_B}}{d E_k},
 \label{X-sec1}
 \eea
 where $E_k$ is the kinetic energy of final electron,
 $\sigma_{{\vec p}_B}$ the cross section
 of the scattering of a neutrino with an electron of
 energy $E_B=m_e-\varepsilon$ and momentum ${\vec p}_B$.
 Eq. (\ref{X-sec1}) says that the neutrino has
 a probability, $|\varphi_l({\vec p}_B)|^2 d^3 p_B$, to scatter with
 an electron carrying momentum ${\vec p}_B$ and the
 cross section is a sum of contributions of the scattering with
 electrons carrying all possible ${\vec p}_B$.

 The energy and
 momentum conservation conditions for $\sigma_{{\vec p}_B}$ are
 \bea
 E_\nu+E_B=E'_\nu+E_e, ~~
 {\vec p}_\nu+{\vec p}_B = {\vec p}_\nu^{\hspace{0.1cm} \prime} +{\vec p}_E,
 \label{conservation}
 \eea
 where $E_B$ is the energy of bound electron as given above,
 $E_e$ the energy of final electron, $E_\nu$
 the energy of initial neutrino, $E'_\nu$ the energy of final neutrino,
 ${\vec p}_\nu$ the momentum of initial neutrino,
 ${\vec p}_\nu^{\hspace{0.1cm} \prime}$ the momentum of final neutrino,
 ${\vec p}_E$ the momentum of final electron.
 Using Eq. (\ref{chiral}) one can find that
 \bea
 \frac{d\sigma_{{\vec p}_B}}{d E_k}
 =\frac{G_F^2}{4\pi^2 |{\vec p}_{\nu B}|} ~\int_0^{2\pi} ~d\phi
 ~\frac{1}{2}\sum_{spin}|{\cal M}|^2 \label{X-sec1a}
 \eea
 where ${\vec p}_{\nu B}={\vec p}_\nu+{\vec p}_B$,
 $\phi$ is the azimuthal angle of ${\vec p}_E$ with respect
 to the axis of ${\vec p}_{\nu B}$ and
 \bea
 \frac{1}{2}\sum_{spin}|{\cal M}|^2=
 (g_V+g_A)^2 p'_\nu\cdot p_E +(g_V-g_A)^2 p_\nu \cdot p_E\frac{E'_\nu}{E_\nu}
 -(g_V^2-g_A^2)\frac{m_e}{E_\nu}p_\nu \cdot p'_\nu .
 \label{X-sec1b}
 \eea 
 $p_\nu, p'_\nu, p_E$ are the four-momenta of initial neutrino,
 final neutrino and final electron respectively.
 In Eq. (\ref{X-sec1b}) an average over spin of the initial electron
 has been performed.

 Using Eq. (\ref{conservation}) one can find that
 \bea
 && p'_\nu \cdot p_E=\frac{1}{2}(E_T^2-|{\vec p}_{\nu B}|^2-m_e^2), \label{condition1} \\
 && 2{\vec p}_{\nu B} \cdot {\vec p}_E=|{\vec p}_{\nu B}|^2+|{\vec p}_E|^2
 -(E_T-E_e)^2  \label{condition2}
 \eea
 where $E_T=E_\nu+E_B$ is the total energy of the scattering
 process. One can see that the projection of ${\vec p}_E$ onto
 the axis of ${\vec p}_{\nu B}$ is fixed by the energy-momentum
 conservation condition but the azimuthal angle $\phi$ is not fixed.
 Using Eq. (\ref{condition2}) one can easily find that 
 \bea
 \frac{1}{2\pi}~\int_0^{2 \pi} ~d\phi~ {\vec p}_E
 = {\vec p}_{\nu B} ~S, \label{condition3}
 \eea
 where 
 \bea
 S= \frac{|{\vec p}_{\nu B}|^2+|{\vec p}_E|^2 -(E_T-E_e)^2}{2 |{\vec p}_{\nu B}|^2}.
 \eea
 Using Eq. (\ref{condition3}) one can find that
 \bea
 \frac{d \sigma_{{\vec p}_B}}{d E_k}=
 \frac{G_F^2}{2\pi |{\vec p}_{\nu B}|} \frac{1}{2}\sum_{spin}|{\tilde M}|^2
 \label{X-sec2}
 \eea
 where
 \bea
 \frac{1}{2}\sum_{spin}|{\tilde M}|^2
&& =\frac{1}{2}(g_V+g_A)^2(E_T^2-|{\vec p}_{\nu B}|^2-m_e^2)
+(g_V-g_A)^2(E_\nu E_e -{\vec p}_\nu\cdot {\vec p}_{\nu B}S) \frac{E'_\nu}{E_\nu} \nnb \\
&& -(g_V^2-g_A^2)\frac{m_e}{E_\nu}[E_\nu(E_T-E_e)-{\vec p}_\nu\cdot {\vec p}_{\nu B}
 (1-S)].
 \label{X-sec2a}
 \eea
 For ${\vec p}_B=0$ and $E_B=m_e$ one can easily show that
 Eq. (\ref{X-sec2}) recovers Eq. (\ref{X-sec0}).

 Using Eq. (\ref{conservation}) one can find that
 the energies of final electron and final neutrino lie in the following range
 \bea
 E_e &&\subset\bigg[\frac{(E_T-|{\vec p}_{\nu B}|)^2+m_e^2}{2(E_T-|{\vec p}_{\nu B}|)},
 \frac{(E_T+|{\vec p}_{\nu B}|)^2+m_e^2}{2(E_T+|{\vec p}_{\nu B}|)}\bigg]
  \label{range1} \\
E'_\nu &&\subset\bigg[\frac{E_T^2-|{\vec p}_{\nu B}|^2-m_e^2}{2(E_T+|{\vec p}_{\nu B}|)},
 \frac{E_T^2-|{\vec p}_{\nu B}|^2-m_e^2}{2(E_T-|{\vec p}_{\nu B}|)} \bigg]. 
 \label{range2}
 \eea

 A number of comments are as follows:

 A) From Eqs. (\ref{condition2}), (\ref{range1}) and (\ref{range2})
 one can read out that not all electrons with all possible
 ${\vec p}_B$ can contribute
 to the cross section. Some ${\vec p}_B$ range is forbidden due to
 kinematical constraint. ${\vec p}_B$ allowed to contribute to
 the scattering process should satisfy
 \bea
 E_T^2 -|{\vec p}_{\nu B}|^2-m_e^2 \ge 0. \label{condtion4}
 \eea
 If $E_T^2 -m_e^2 \le 0$, no ${\vec p}_B$ range can contribute
 to the kinematically allowed range and the process is forbidden. This
 is just the threshold condition for the process to happen,
 i.e. $E_\nu > \varepsilon$.

 B) From Eqs. (\ref{range1}) and (\ref{range2}) one can read out that
 the energy ranges of the final particles depend on $|{\vec p}_{\nu B}|$.
 The total width of the range of electron energy is 
 \bea
 \Delta E_e= E_e|_{max}-E_e|_{min}=\frac{E_T^2 -|{\vec p}_{\nu B}|^2-m_e^2}
 {E_T^2 -|{\vec p}_{\nu B}|^2} |{\vec p}_{\nu B}|.
 \label{range3}
  \eea

 C) One can check in Eq. (\ref{range1}) that the minimum
 of the energy of final electron is larger than or equals to $m_e$. To
 allow the energy range to reach $m_e$ it requires
 \bea
 E_T-|{\vec p}_{\nu B}|=m_e.
 \eea
 This requires that ${\vec p}_B$ lies
 in a very narrow range. Hence, the differential cross
 section for $E_k\approx 0$ range is suppressed by
 a factor arising from the momentum distribution of ${\vec p}_B$:
 $|\varphi({\vec p}_B)|^2 \delta^3 p_B$.

 D) From Eq. (\ref{range1}) one can show that $E_e|_{\max}$
 increases with $|{\vec p}_{\nu B}|$. Using Eq. (\ref{condtion4})
 one can find out the maximum energy
 of final electron in the scattering process:
 \bea
 E_e \le \frac{1}{2}
\bigg(E_T+\sqrt{E_T^2-m_e^2}+\frac{m_e^2}{E_T+\sqrt{E_T^2-m_e^2}} \bigg)
 =E_T.
\label{range4}
 \eea
As a comparison, we can find in Eq. (\ref{range0})
that for the scattering with free electron at rest the maximum energy 
of final electron is $\frac{1}{2}[m_e+2E_\nu+m_e^2/(m_e+2E_\nu)]$.
For $E_\nu \gg \varepsilon$ we can easily figure out that the 
energy of final electron extends beyond the range allowed
by the scattering with free electron at rest.

E) Eqs. (\ref{X-sec1}) and (\ref{X-sec2}) appear to be singular 
for $|{\vec p}_{\nu B}|\to 0$.
This is an artificial singularity. As can be seen in Eq. (\ref{range3}),
the energy range of final electron shrinks also as $|{\vec p}_{\nu B}|\to 0$.
This cancels the singular term in Eqs. (\ref{X-sec1}) and (\ref{X-sec2}) in
the integrated cross-section. Furthermore, one can see
that $|{\vec p}_{\nu B}| \approx 0$ corresponds to a very narrow
range of ${\vec p}_B$. In numerical calculation, the contribution
to differential cross section of electrons in this narrow range
can be easily controlled by taking it as $|\varphi({\vec p}_B)|^2
~\delta^3 p_B ~\Delta E_e/|{\vec p}_{\nu B}| 
 ~\frac{1}{2}G_F^2 |{\tilde M}|^2/(2\pi)$.
Using Eq. (\ref{range3}), one can see that it is suppressed by the factor
$|\varphi({\vec p}_B)|^2 ~\delta^3 p_B$ and in practice one can
 neglect it by taking a small enough range of $\delta^3 p_B$.

{\bf Numerical result and discussion}
\begin{figure}[tb]
\begin{center}
\begin{tabular}{cc}
\includegraphics[scale=1,width=7.5cm]{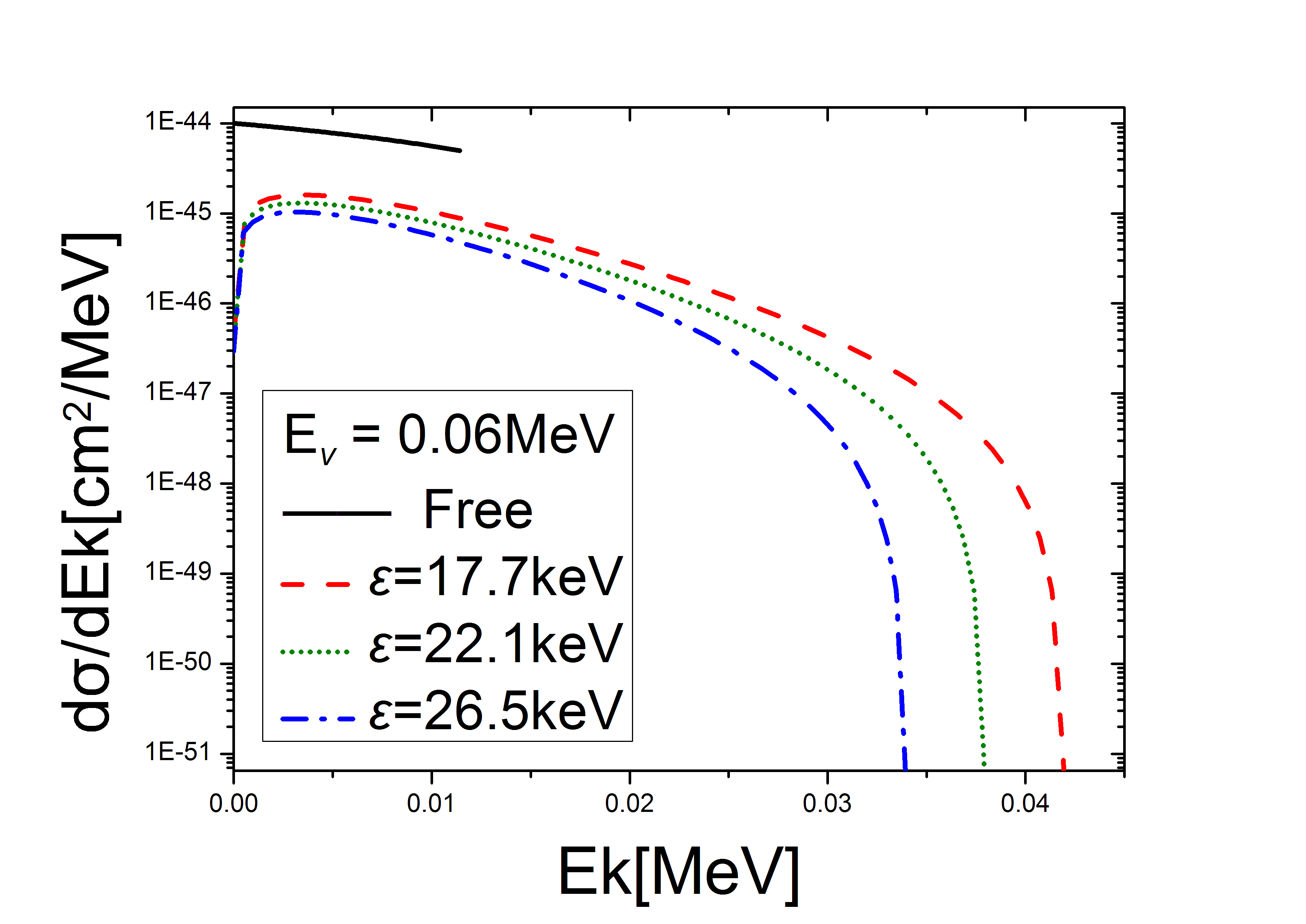}
\includegraphics[scale=1,width=7.5cm]{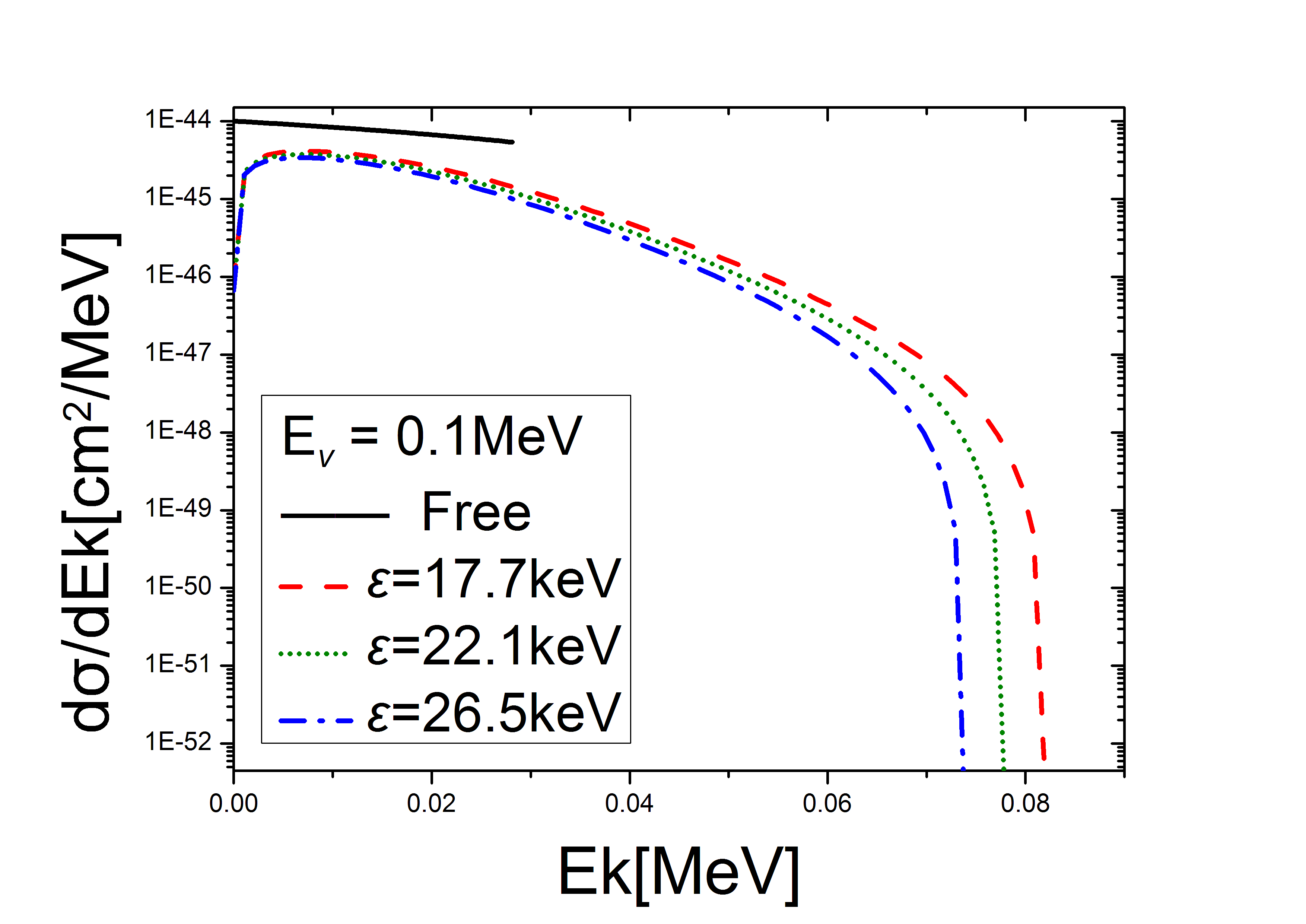}
\\
\includegraphics[scale=1,width=7.5cm]{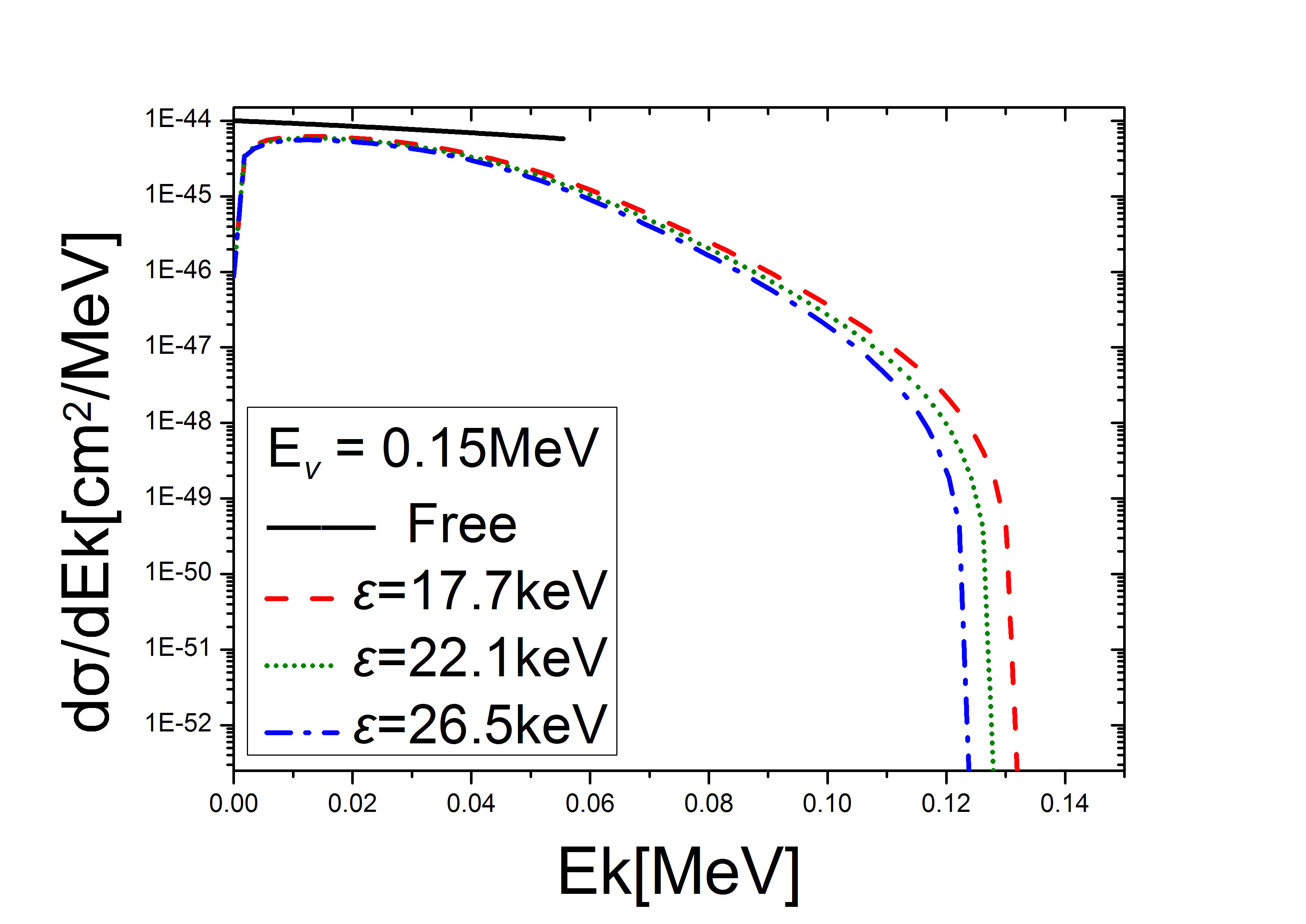}
\includegraphics[scale=1,width=7.5cm]{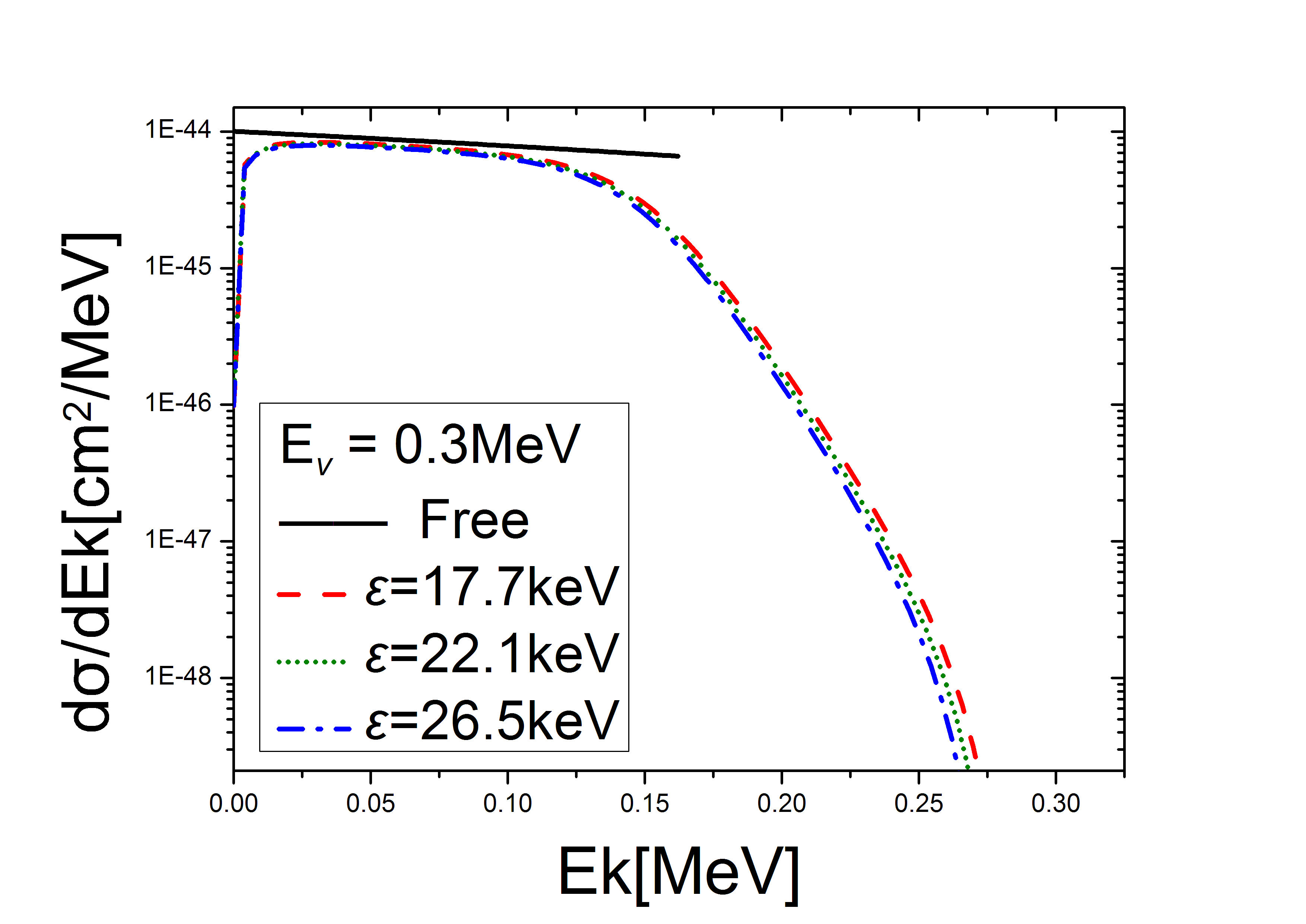}
\\
\includegraphics[scale=1,width=7.5cm]{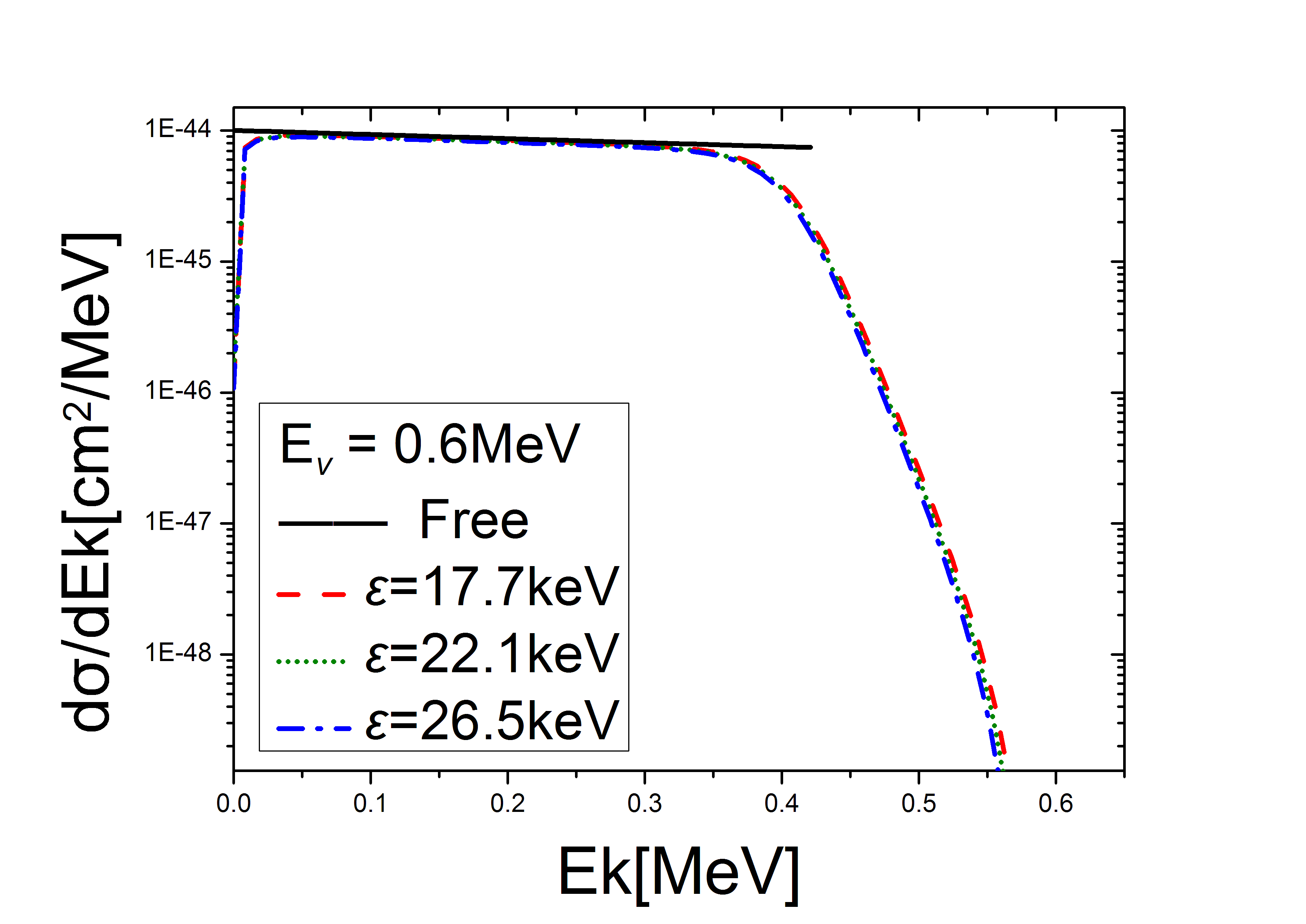}
\includegraphics[scale=1,width=7.5cm]{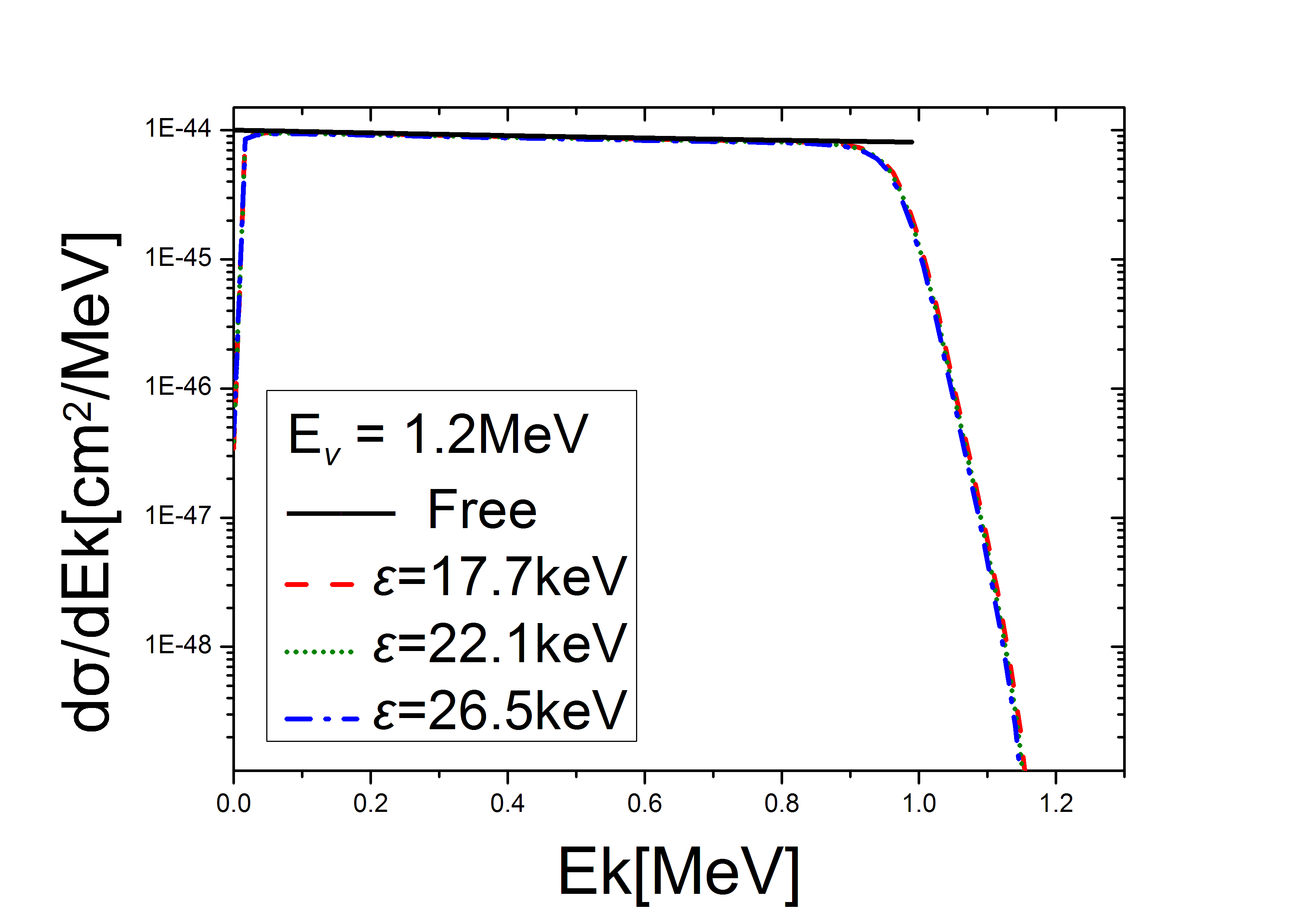}
\end{tabular}
\end{center}
\caption{Differential cross section of the scattering of an electron neutrino 
with bound electron in K shell of Ru atom. The black solid line is
for the scattering with electron at rest with $E_k$ in the kinematically
allowed range $[0, E_\nu/(1+m_e/(2E_\nu))]$. Three colored lines are for
$\varepsilon=17.7, 22.1,26.5$ keV respectively. }
\label{figure1}
\end{figure}

\begin{figure}[tb]
\begin{center}
\begin{tabular}{cc}
\includegraphics[scale=1,width=7.5cm]{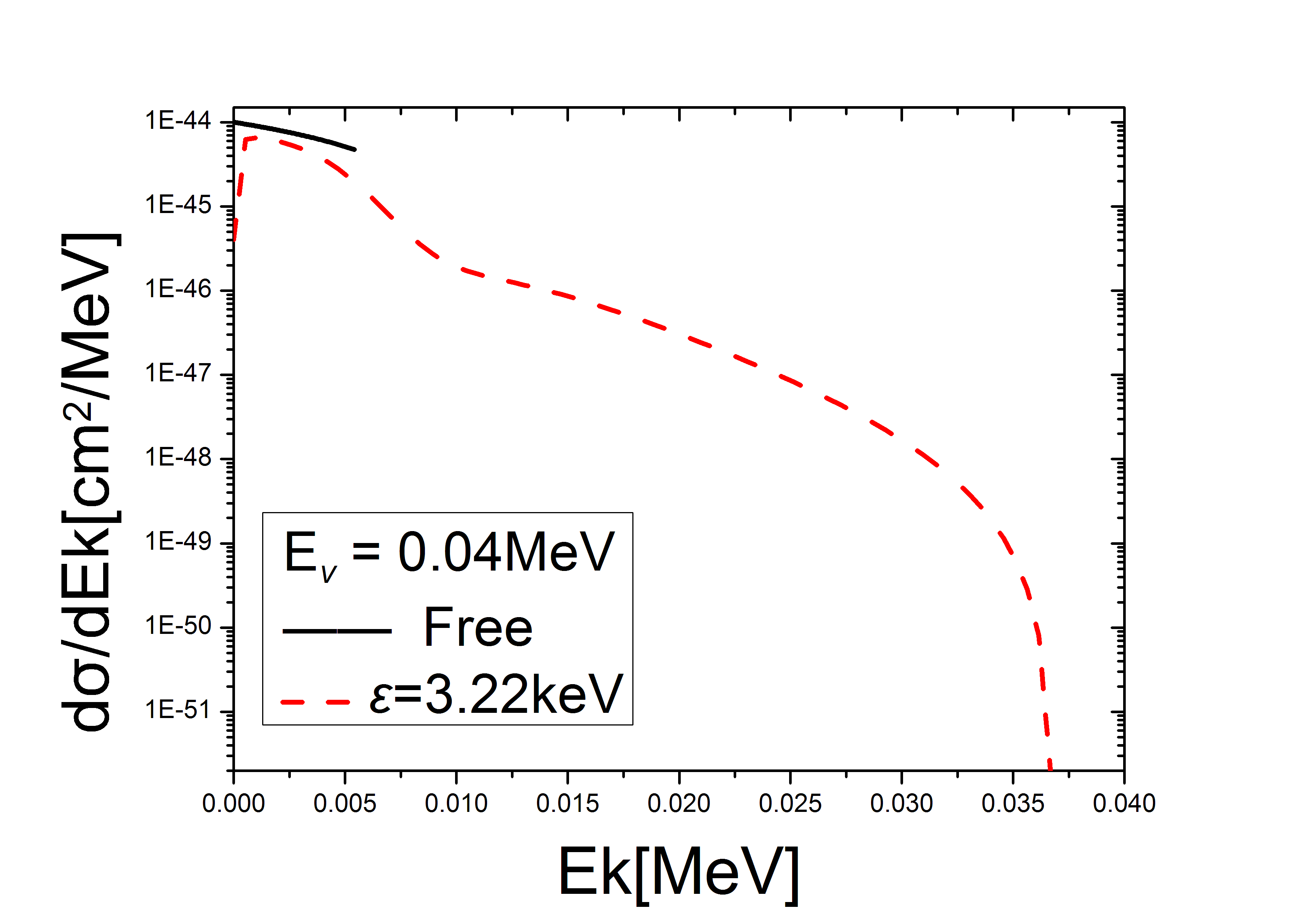}
\includegraphics[scale=1,width=7.5cm]{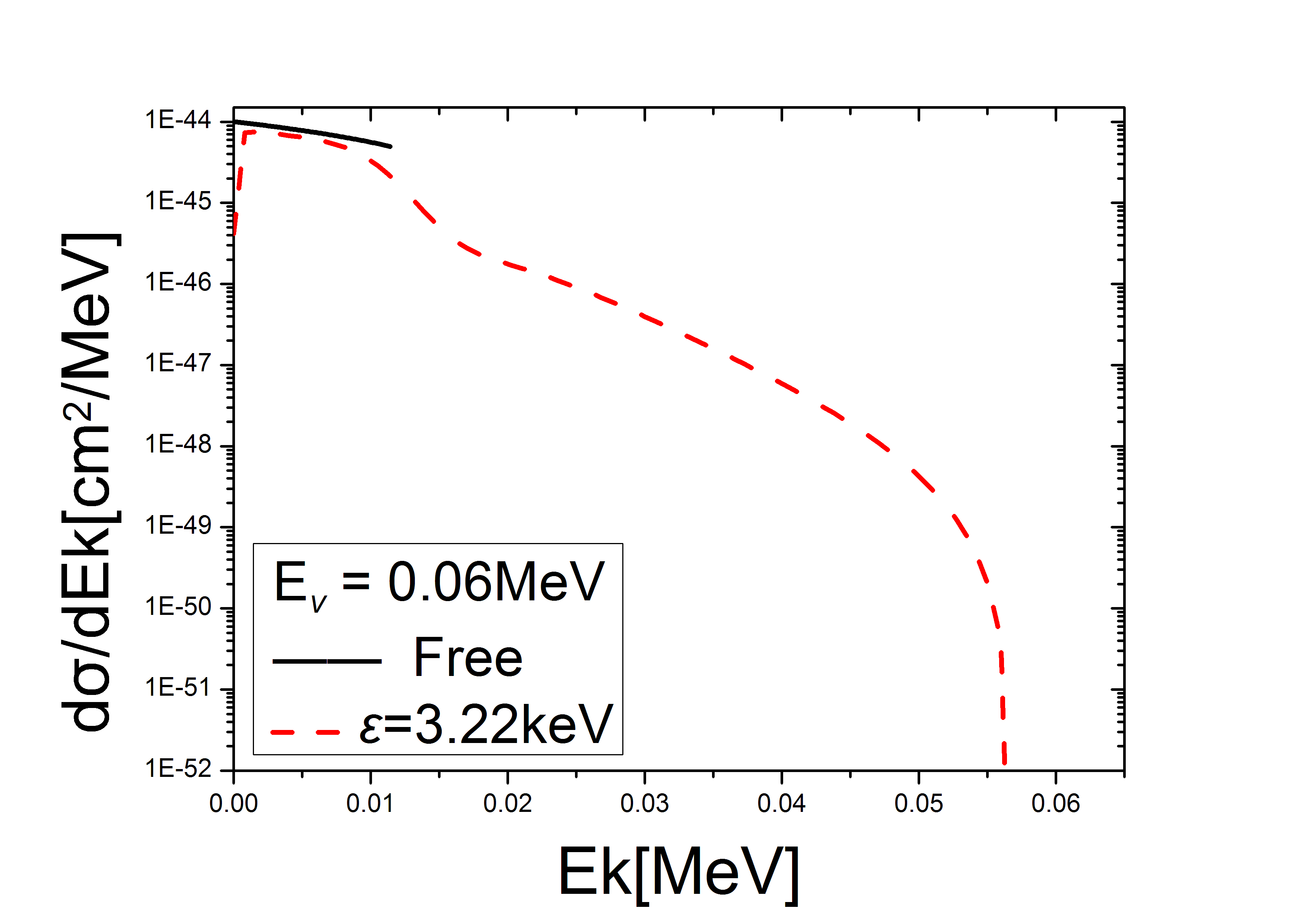}
\\
\includegraphics[scale=1,width=7.5cm]{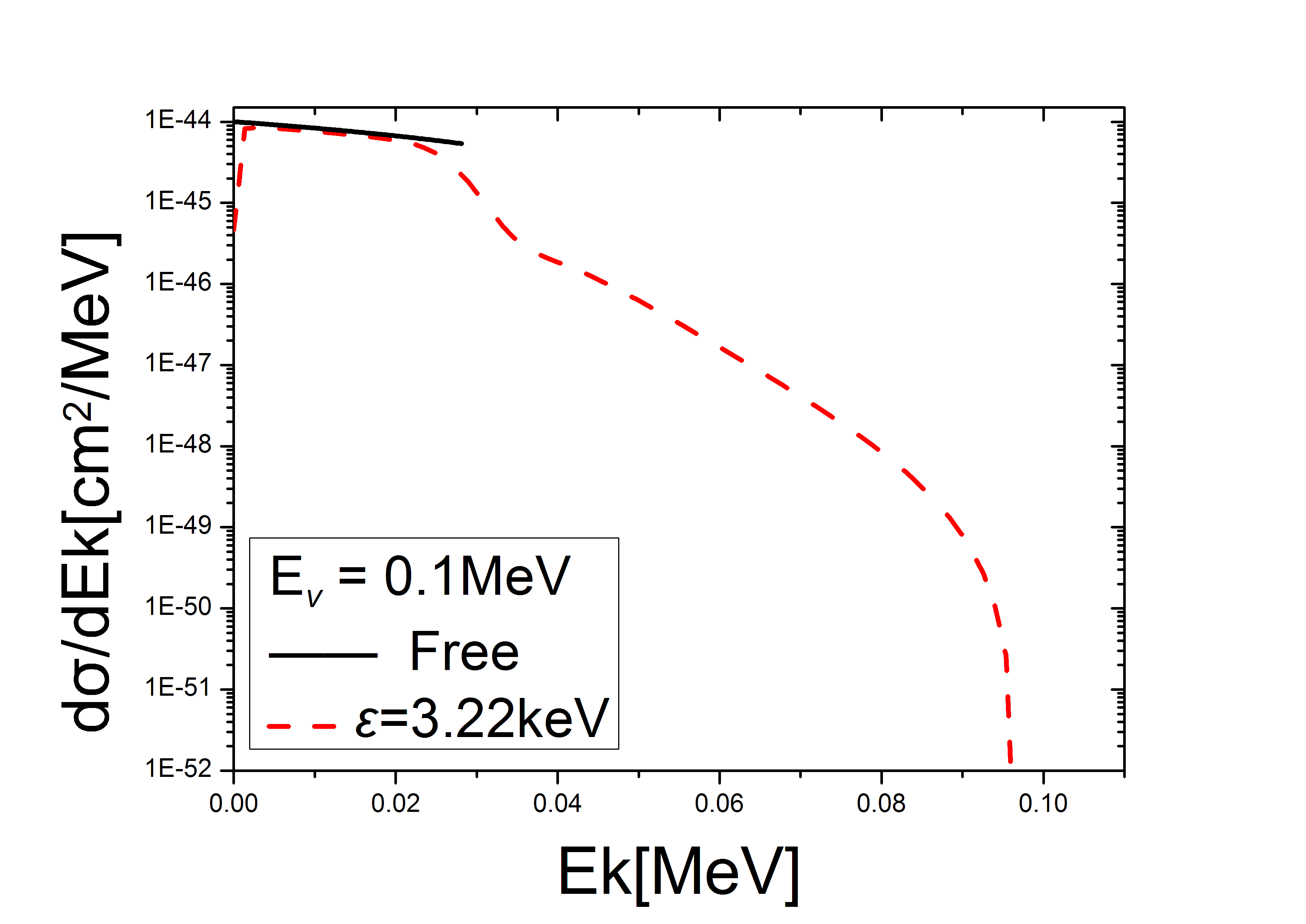}
\includegraphics[scale=1,width=7.5cm]{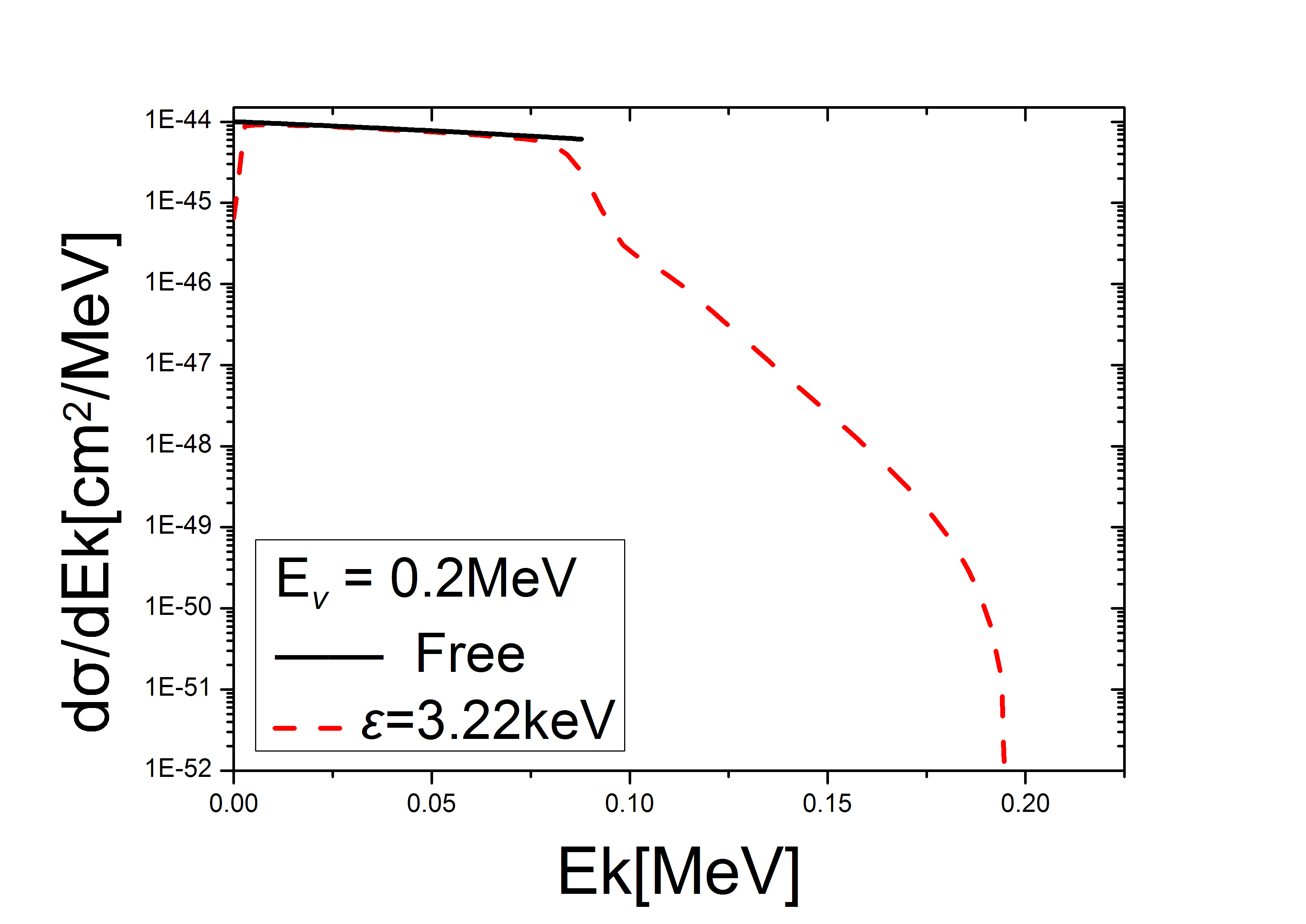}
\\
\includegraphics[scale=1,width=7.5cm]{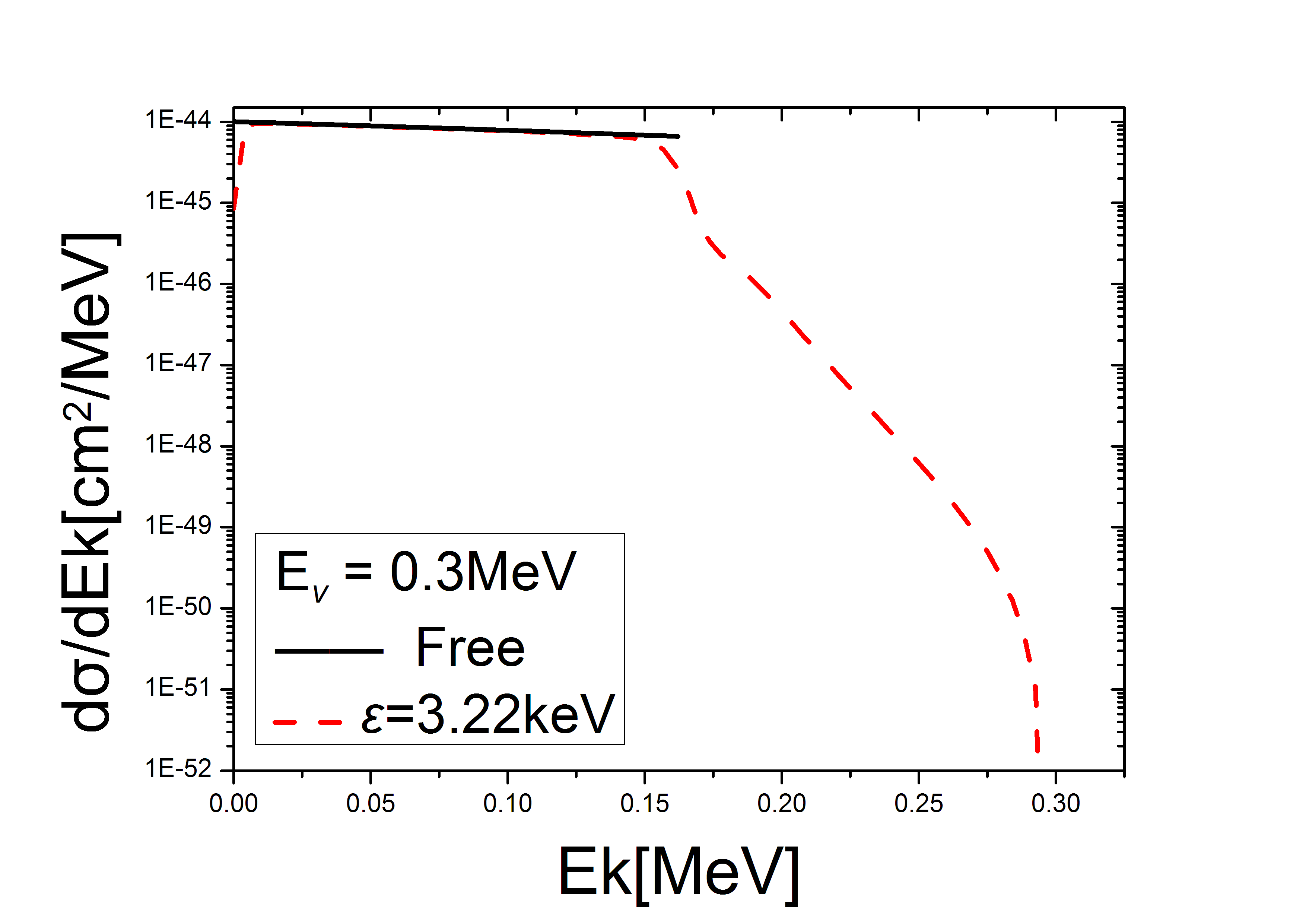}
\includegraphics[scale=1,width=7.5cm]{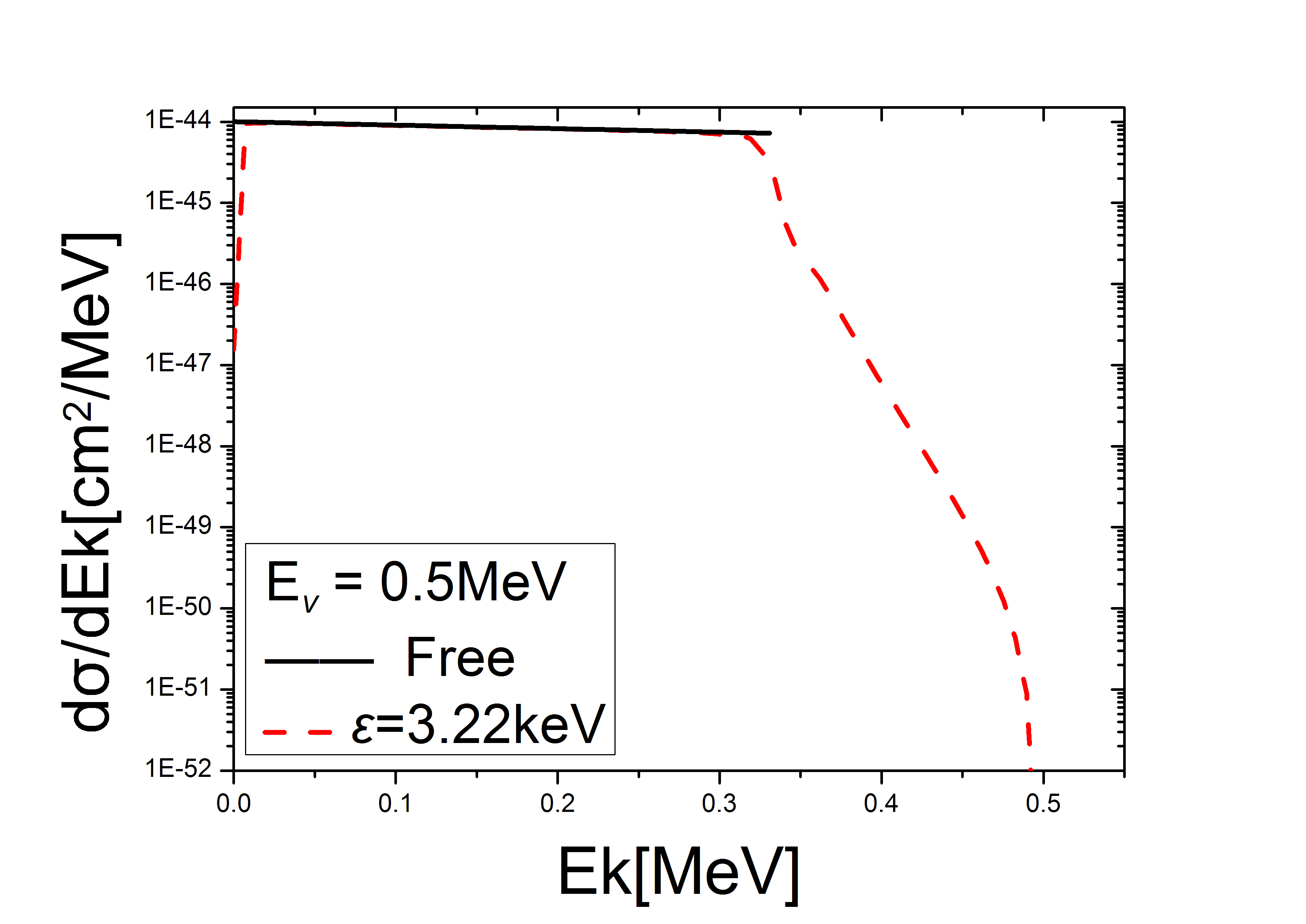}
\end{tabular}
\end{center}
\caption{Differential cross section of the scattering of an electron neutrino 
with bound electron in $2s$ state of Ru atom. The black solid line is
for the scattering with electron at rest with $E_k$ in the kinematically 
allowed range $[0, E_\nu/(1+m_e/(2E_\nu))]$. The binding energy is chosen
as $\varepsilon=3.22$ keV. }
\label{figure2}
\end{figure}

\begin{figure}[tb]
\begin{center}
\begin{tabular}{cc}
\includegraphics[scale=1,width=7.5cm]{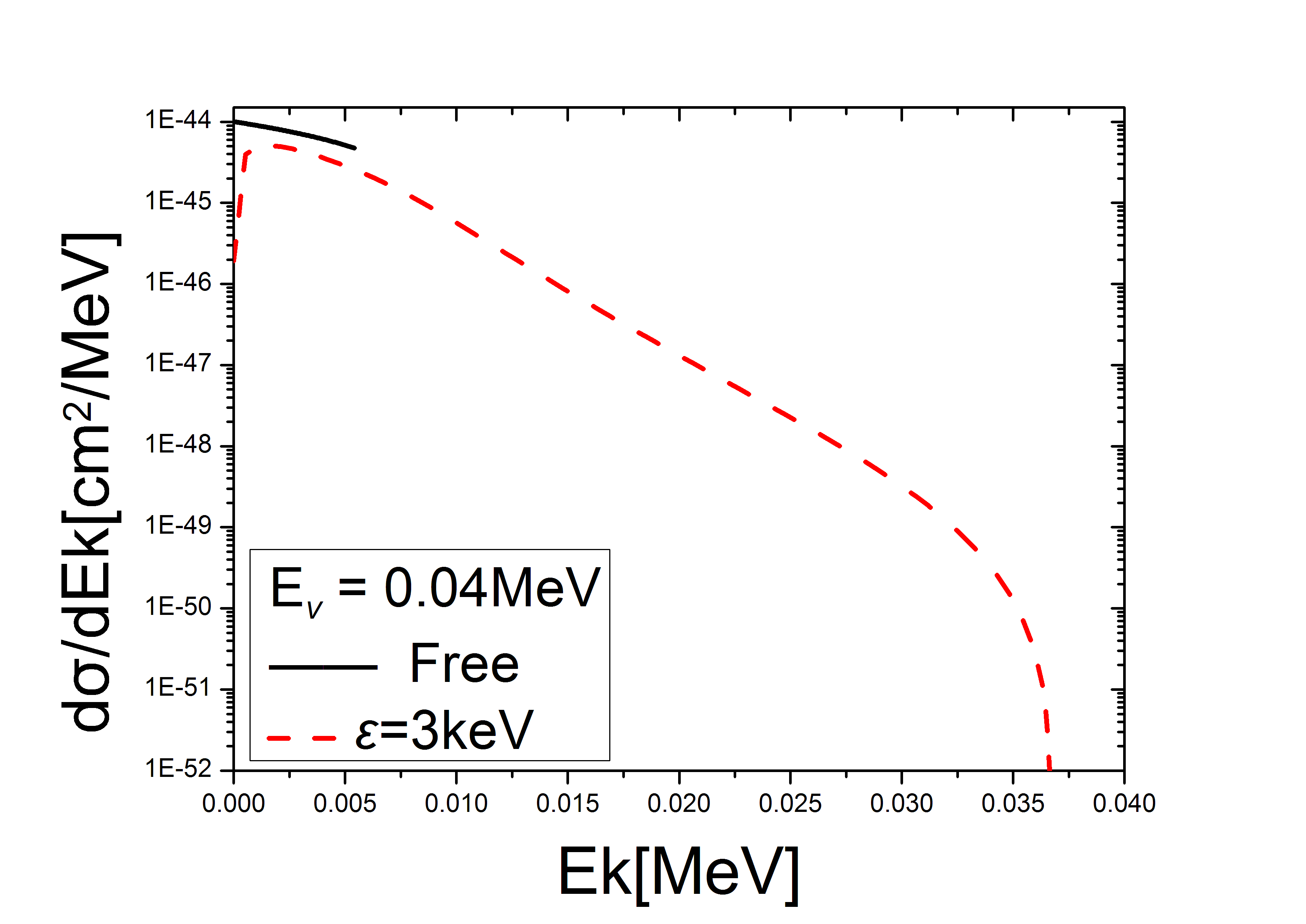}
\includegraphics[scale=1,width=7.5cm]{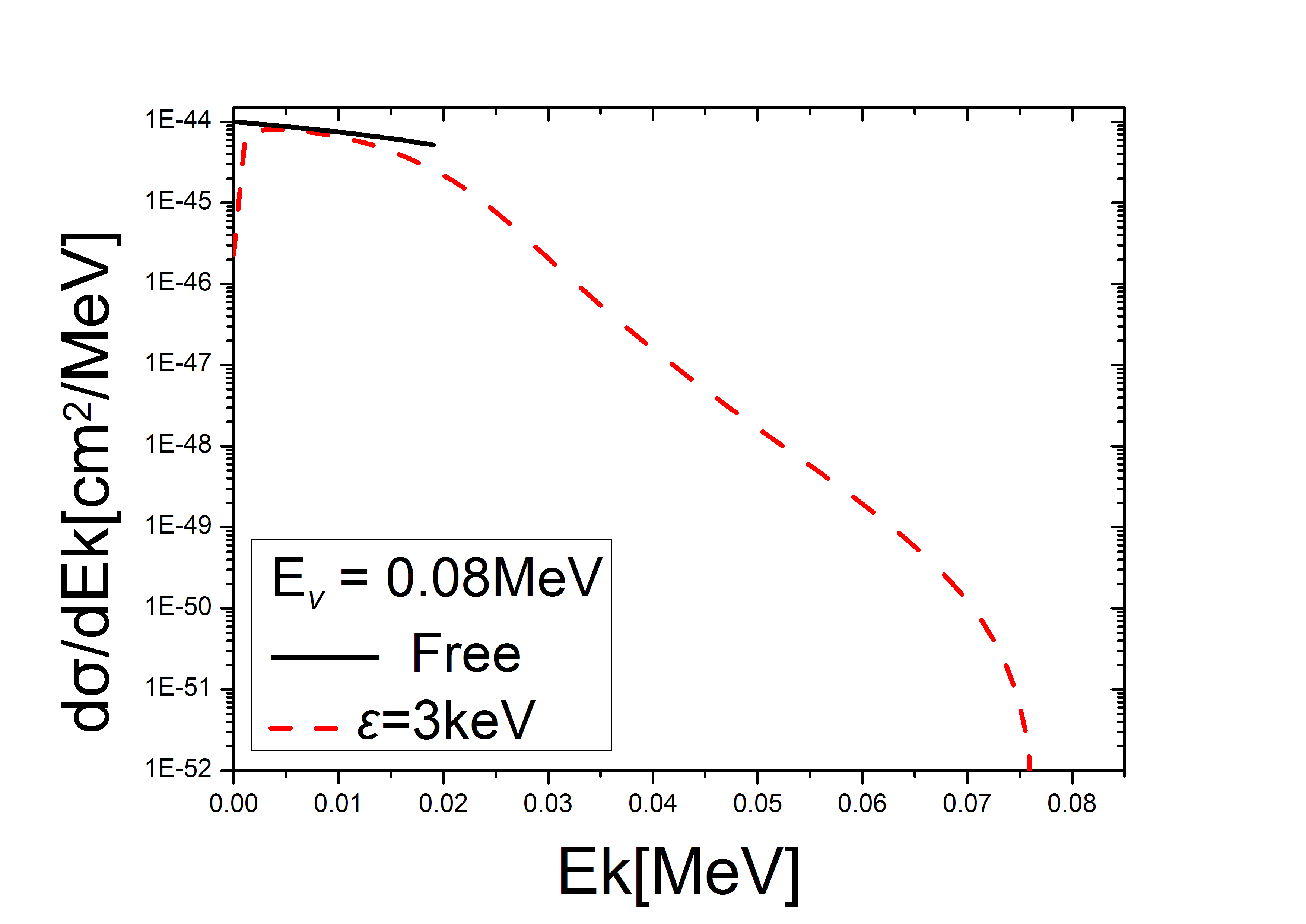}
\\
\includegraphics[scale=1,width=7.5cm]{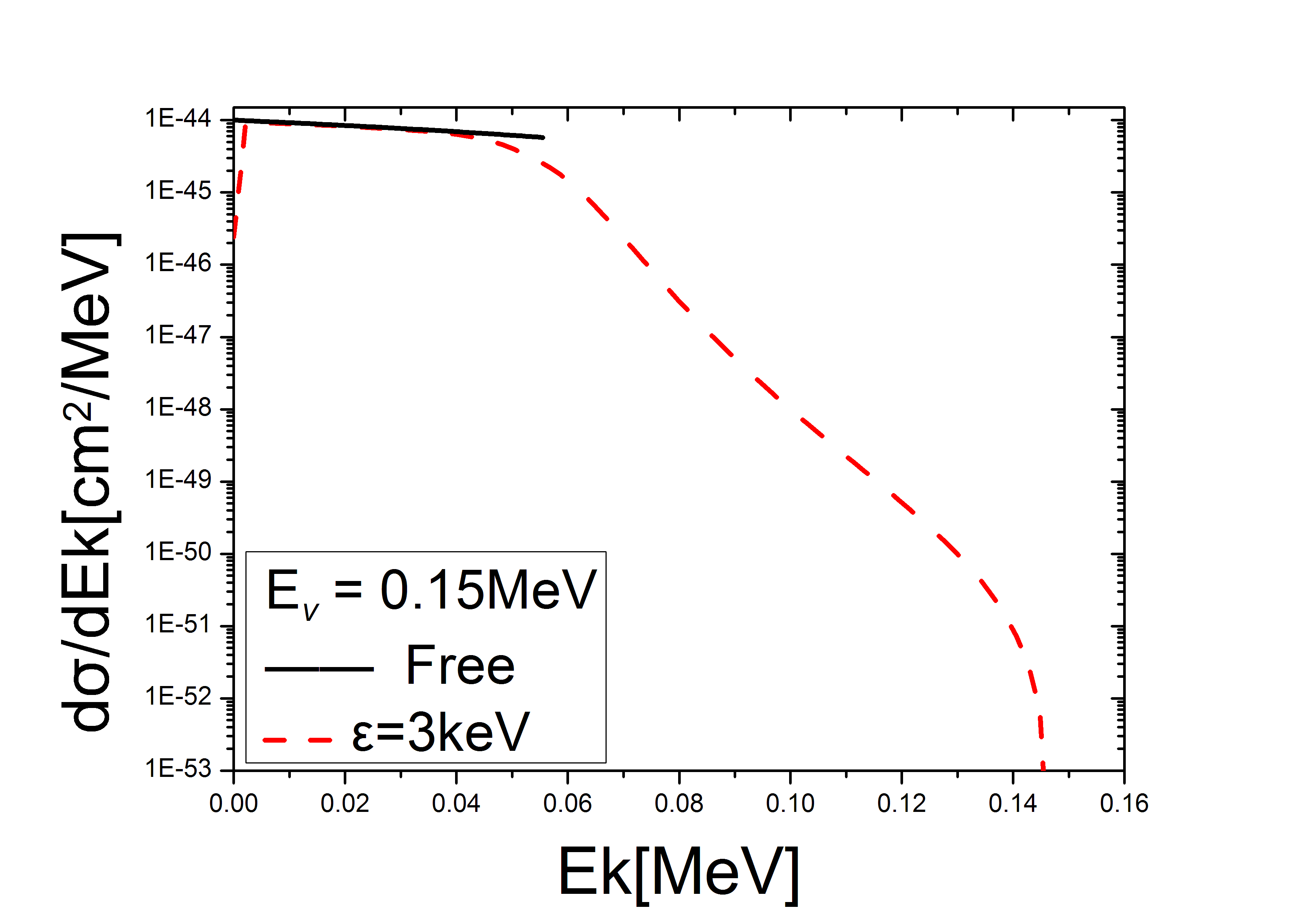}
\includegraphics[scale=1,width=7.5cm]{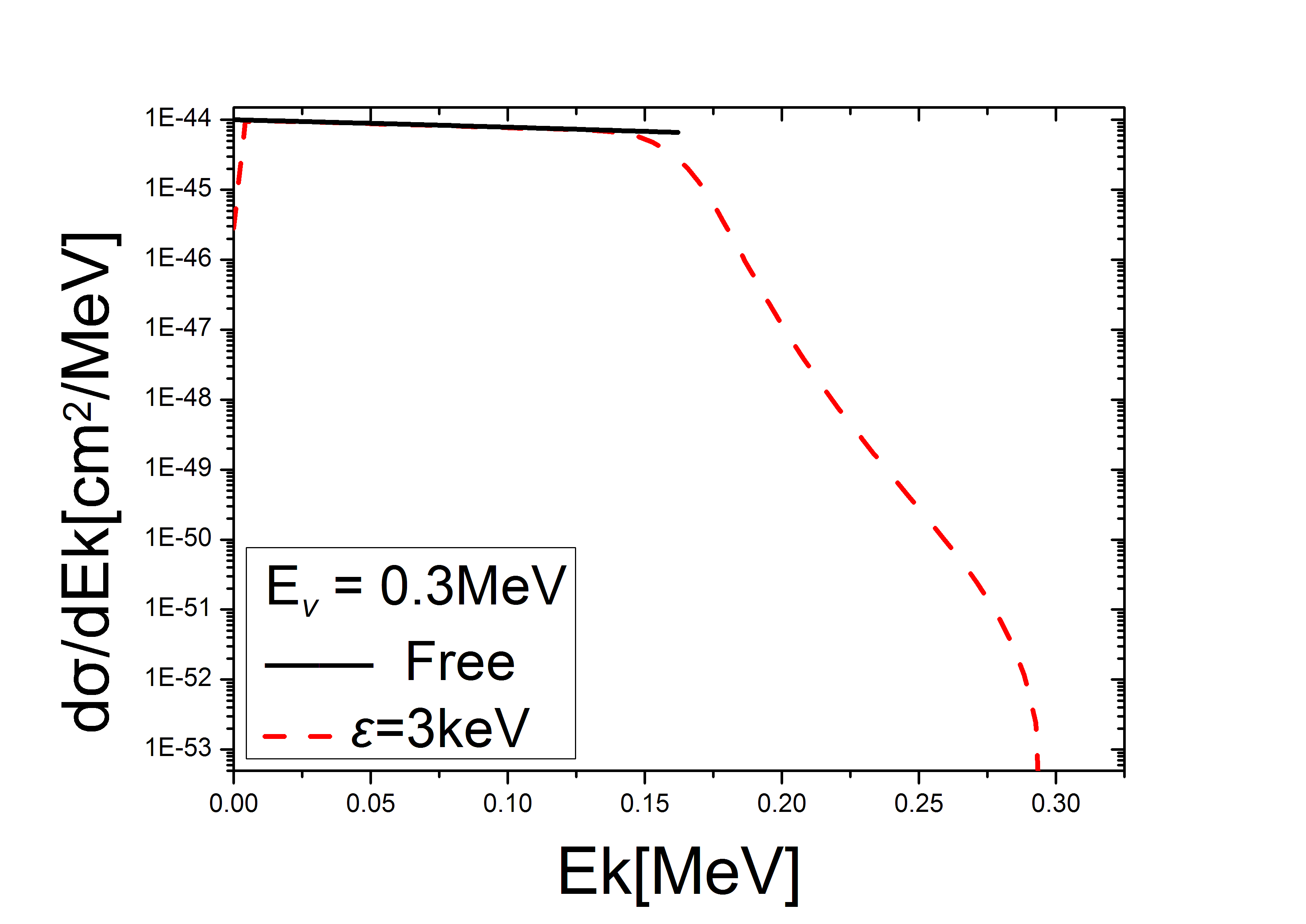}
\\
\includegraphics[scale=1,width=7.5cm]{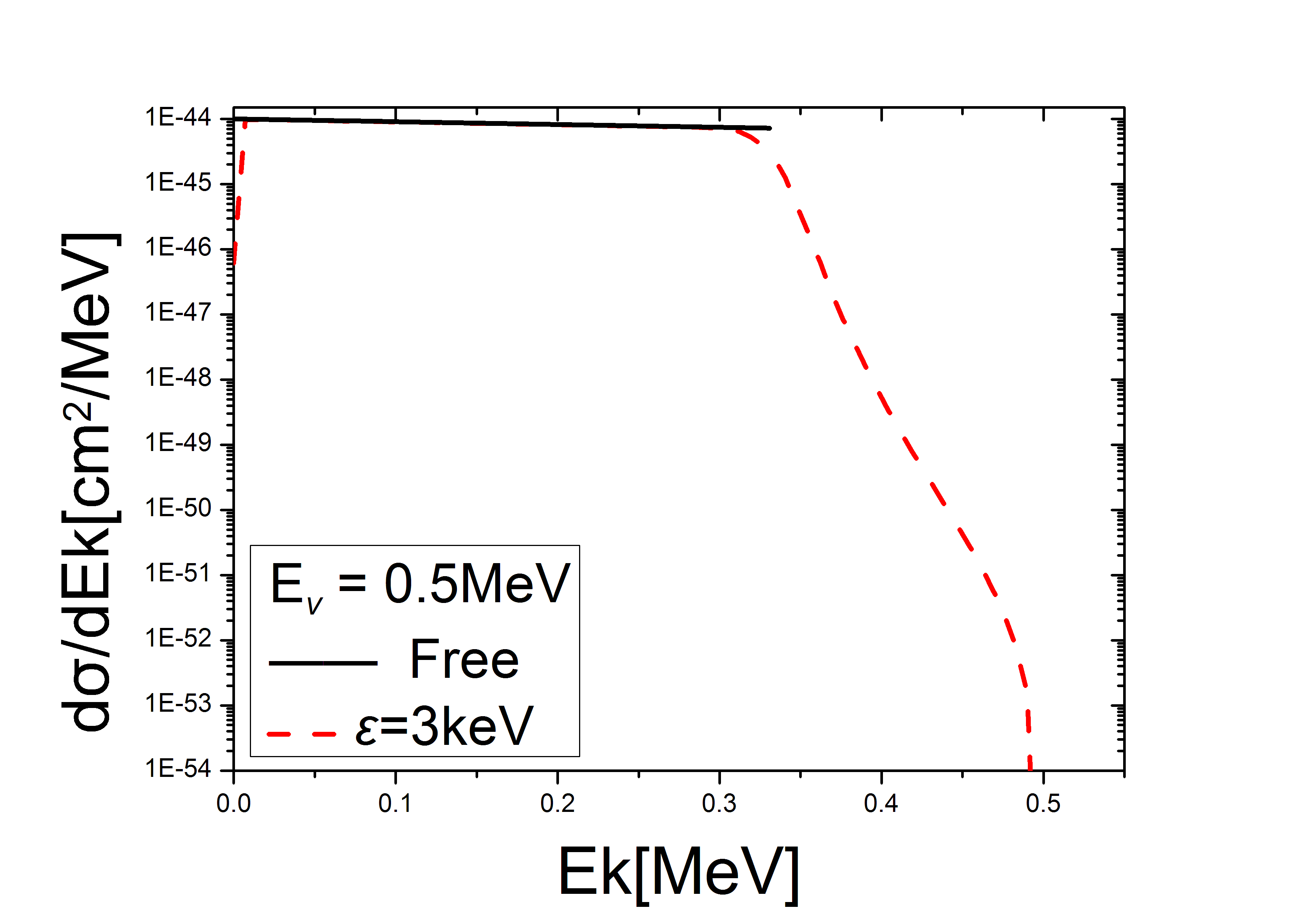}
\includegraphics[scale=1,width=7.5cm]{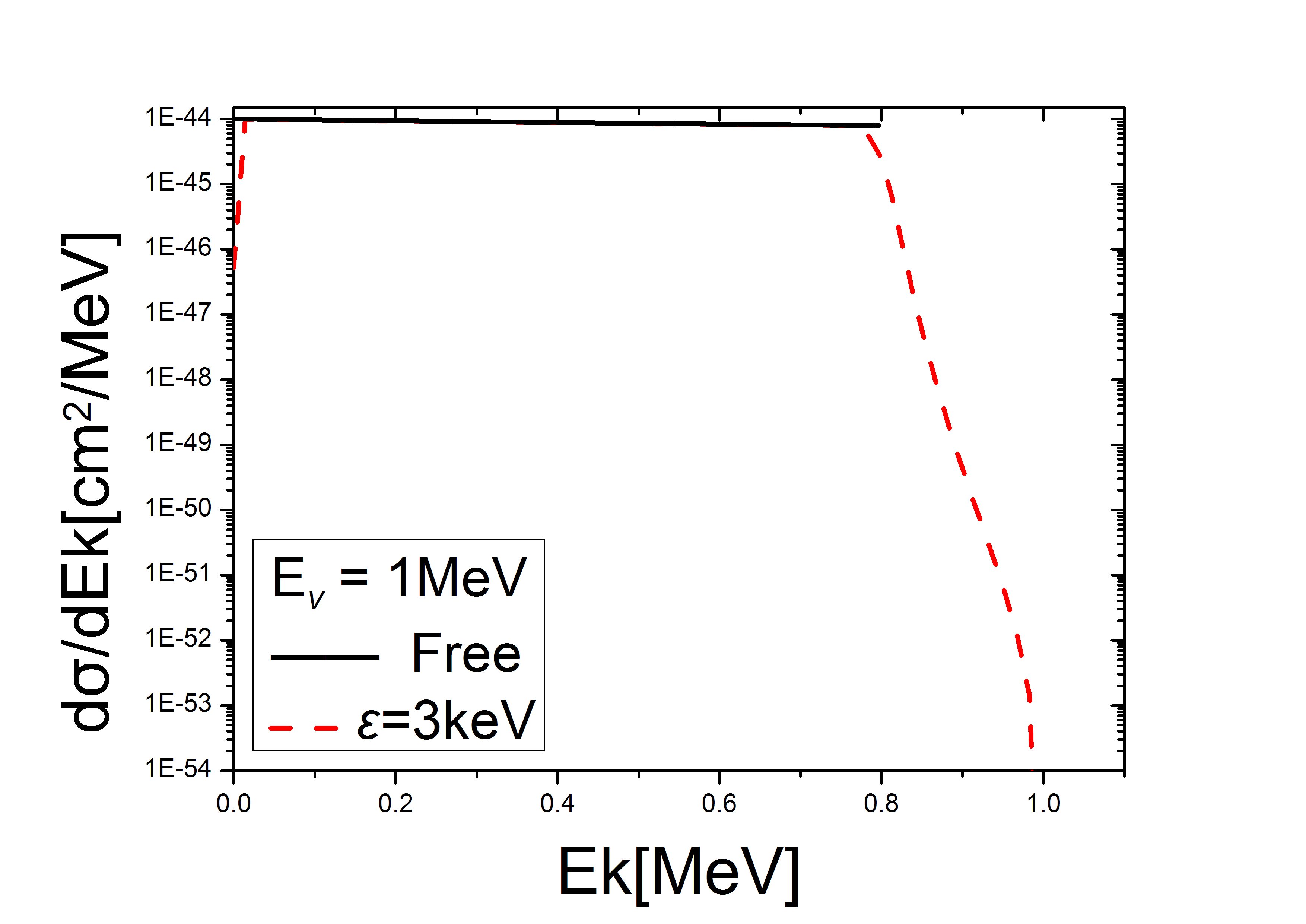}
\end{tabular}
\end{center}
\caption{Differential cross section of the scattering of an electron neutrino 
with bound electron in $2p$ state of Ru atom.The black solid line is
for the scattering with electron at rest with $E_k$ lies in the kinematically
allowed range $[0, E_\nu/(1+m_e/(2E_\nu))]$. The binding energy is chosen
as $\varepsilon=3$ keV. }
\label{figure3}
\end{figure}

In Fig. \ref{figure1} we show the differential cross section 
for the scattering of an electron neutrino with electron in K shell of Ru
atom. One can see that at $E_k\approx 0$ tail the differential
cross section drops down sharply. As a comparison, one can
see that the cross section of the scattering with free electron
at rest remains constant for $E_k=0$. As noted in comment C)
shown above, this is because the initial electron has a distribution
of momentum and the probability of finding electron in
the required momentum range, i.e. the range for reaching
$E_k\approx 0$, is small.

In Fig. \ref{figure1} we can see that the scattering of a neutrino with 
a bound electron has an $E_k$ spectrum wider than that of the scattering 
with free electron
at rest. As shown in Eq. (\ref{range0}), the kinetic energy of the final
electron in the scattering with free electron at rest lies
in a limited range with a maximum value which can be
much smaller than $E_\nu$ when $E_\nu < m_e$. In contrast,
the final electron is allowed to get a kinetic energy as large
as $E_\nu-\varepsilon$ for the scattering with bound electron, as
noted in comment D) above.
One can find that for $E_\nu \lsim 0.15$ MeV the difference between 
the allowed energy ranges 
of the scattering with free electron and the scattering 
with bound electron is very large.
For $E_\nu >m_e$, the difference becomes negligible.

One can see in Fig. \ref{figure1} that for large $E_\nu$
the cross section of the scattering with bound electron 
becomes close to that of the scattering with free
electron at rest. In particular, for $E_\nu \gsim 0.3$ MeV
the cross section of the scattering with bound electron
agrees well with that of the free electron case in the kinematically
allowed region of $E_k$ in the free electron case. 
Although
there is still a tail beyond the range allowed by the
scattering with free electron at rest, the differential cross section
in tail region is suppressed by several orders of magnitude. 

\begin{figure}[tb]
\begin{center}
\begin{tabular}{cc}
\includegraphics[scale=1,width=10.5cm]{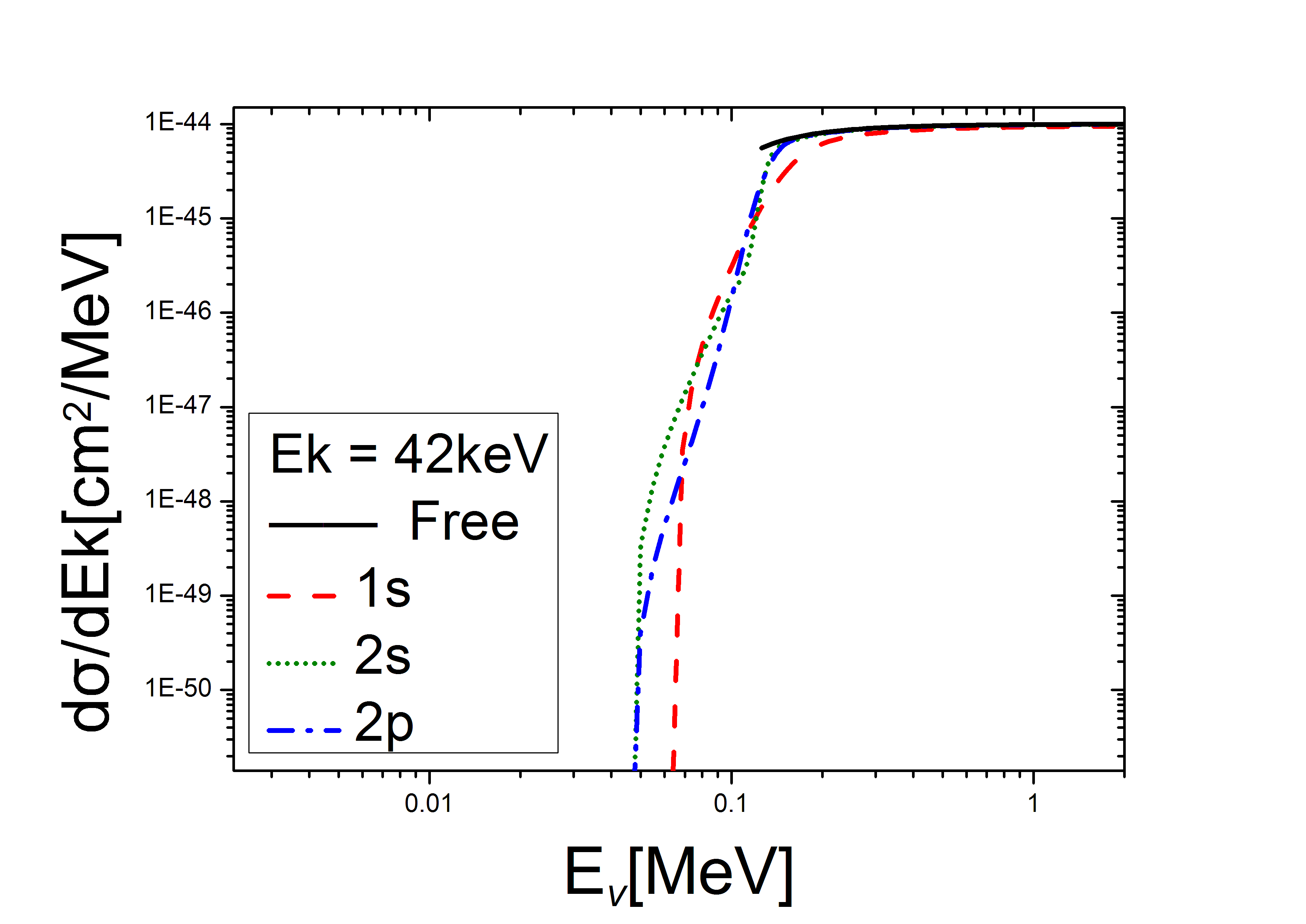}
\end{tabular}
\end{center}
\caption{Differential cross section versus $E_\nu$, the energy of the initial
neutrino, for the scattering of an electron neutrino 
with bound electron in Ru atom. The binding energies for
1s, 2s and 2p states are chosen the same as in Figs. \ref{figure1},
\ref{figure2} and \ref{figure3} saperately.The energy of the final electron
is fixed as $E_k=42$ keV.}
\label{figure4}
\end{figure}

As a comparison we plot three lines in Fig. \ref{figure1}
for $\varepsilon=22.1$ keV and $\pm 20\%$ variation of this
value of binding energy. One can see that $20\%$ variation does not lead to
large difference for the scattering with $E_\nu > 0.1$ MeV.
It has impact to the scattering with $E_\nu= 0.06$ MeV.
In particular, there are visible differences close to the end of the tails
in the plot for $E_\nu=0.06$ MeV. This corresponds to
the change of the threshold for the production of a final electron
with a particular energy. It's not a surprise that this threshold
would depend on value of $\varepsilon$. 

In Fig. \ref{figure2} we plot the differential cross section
of the scattering of an electron neutrino with electron in $2s$ state.
We can find phenomena similar to that discussed above for the scattering
with electron in $1s$ state. A major difference is that the lines
for the scattering with electron in $2s$ state get close to that of
the free electron case at smaller $E_\nu$ compared to the lines of
the scattering with electron in $1s$ state. One can see in 
Fig. \ref{figure2} that for $E_\nu \gsim 0.06$ MeV the differential
cross section of the scattering with bound electron is already
quite close to that of the scattering with free electron in the
energy range, $[0,E_\nu/(1+m_e/(2E_\nu))]$, 
allowed by the scattering with free electron.

In Fig. \ref{figure3} we plot the differential cross section
of the scattering of an electron neutrino with electron in $2p$ state
of Ru atom. The plot is given for the cross section evaluated
for the average momentum distribution
$|\varphi_{2p}|^2$ of electrons in $2p$ state:
\bea
 \frac{d \sigma_{2p}}{d E_k}= \int~d^3p_B~
 |\varphi_{2p}({\vec p}_B)|^2 \frac{d\sigma_{{\vec p}_B}}{d E_k},
 \label{X-sec3}
 \eea
where
\bea
|\varphi_{2p}({\vec p}_B)|^2=\frac{1}{3}
(|\varphi_{2p0}({\vec p}_B)|^2+|\varphi_{2p+1}({\vec p}_B)|^2
+|\varphi_{2p-1}({\vec p}_B)|^2).
\label{distrib2p}
\eea 
For convenience we choose a universal binding energy $\varepsilon_L=3$ keV 
for $\varphi_{2p0}$ and $\varphi_{2p\pm1}$. 
We note that $2p_{1 \over 2}$ and $2p_{3 \over 2}$ states
correspond to linear combinations of $\varphi_{2p0}$
 and $\varphi_{2p\pm1}$ together with spinors. 
An advantage of computing Eq. (\ref{X-sec3}) is that
it's invariant after recombination of wavefunctions $\varphi_{2p0}$
and $\varphi_{2p\pm 1}$.
In particular, computation using $2p_{1\over 2}$ and $2p_{3\over 2}$
wavefunctions lead to the same result as in Eq. (\ref{distrib2p})
after averaging contributions of all electrons in
$2p_{1\over 2}$ and $2p_{3\over 2}$ states.

In Fig. \ref{figure3} we can find phenomena similar to that
discussed for the scattering with electron in $2s$ state.
Similarly, we can find in
Fig. \ref{figure3} that for $E_\nu \gsim 0.06$ MeV the differential
cross section of the scattering with bound electron is already
quite close to that of the scattering with free electron in the
energy range, $[0, E_\nu/(1+m_e/(2E_\nu))]$,
allowed by the scattering with free electron at rest.

\begin{figure}[tb]
\begin{center}
\begin{tabular}{cc}
\includegraphics[scale=1,width=10.5cm]{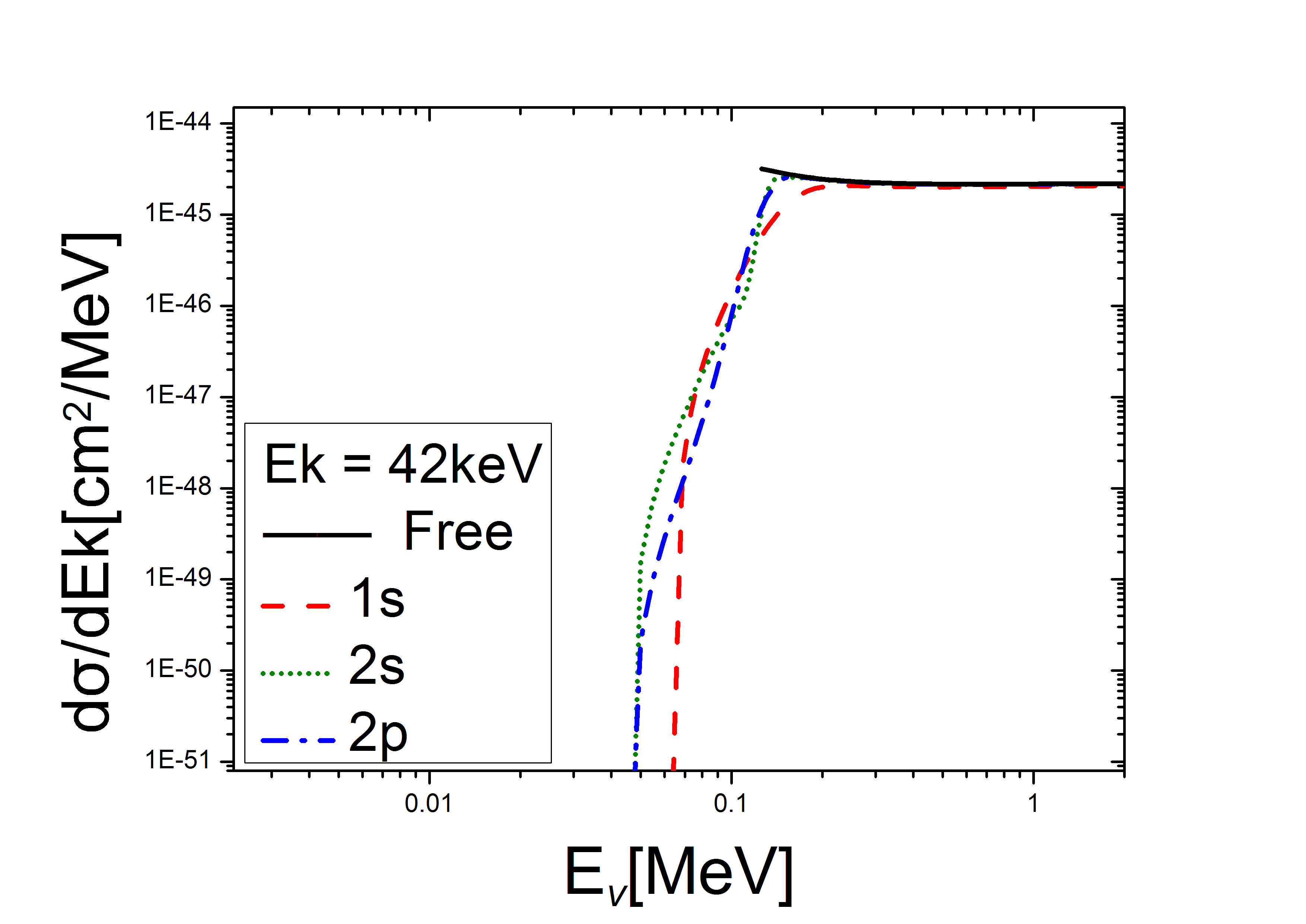}
\end{tabular}
\end{center}
\caption{Differential cross section versus $E_\nu$, the energy of the initial
neutrino, for the scattering of a muon neutrino 
with bound electron in Ru atom. The binding energies for
1s, 2s and 2p states are chosen the same as in Figs. \ref{figure1},
\ref{figure2} and \ref{figure3} saperately. The energy of the final electron
is fixed as $E_k=42$ keV.}
\label{figure5}
\end{figure}

We have seen in Figs. \ref{figure1},\ref{figure2},\ref{figure3} that
the final electron has an energy range wider than that
for the scattering with free electron at rest.
In particular, one can show that
neutrinos with energy $E_\nu < \frac{1}{2}(E_k
+\sqrt{E_k^2+2 E_k m_e})$ do not contribute to the differential
cross section, Eq. (\ref{X-sec0}), for a fixed $E_k >0$.
If considering electron events of a particular kinetic
energy $E_k$, some low energy neutrinos in the initial spectrum
would be found not to contribute to this type of events
when using the cross section of the scattering with free electron
at rest. According to the above discussion, these low
energy neutrinos can indeed contribute to this type
of events when using the cross section of the scattering with
bound electron.

In Fig. \ref{figure4} we plot the differential cross section
for $E_k=42$ keV versus the initial $E_\nu$. 
One can see in Fig. \ref{figure4}
that for the scattering of electron neutrino
with free electron the cross section
starts to be non-zero from a non-zero value of $E_\nu$
which is $(E_k+\sqrt{E_k^2+2m_e E_k})/2$ as discussed above.
For the scattering with bound electron, the cross section
starts to be non-zero from a much smaller value of $E_\nu$ 
than that of the scattering with free electron at rest.
For $E_\nu$ close to the threshold value, the 
differential cross section of the scattering with bound electrons
can start from a value infinitely close to zero. 
One can see that the lines for the scattering with electrons
in 2s or 2p states are very close to the line for the scattering
with free electron at rest in the kinematically allowed 
energy range of the scattering with free electron. 
Outside this allowed energy range,
the cross sections of the scattering with
electrons in 2s or 2p states decrease sharply for orders of magnitude.
This means that the bound state features of 2s and 2p states 
just give small corrections to the scattering process.
For comparison one can see that the line for the scattering with
electron in 1s state is not very close to the line for the free electron
case. There are visible differences between the line for the
free electron case and the line for electron in 1s state up 
to $E_\nu \sim 0.2-0.3$ MeV.
In Fig. \ref{figure5} we give a plot similar to Fig.\ref{figure4}
but for the scattering of muon neutrino with bound electrons.
One can see phenomena similar to that in Fig. \ref{figure4}.

In Figs. \ref{figure4} and \ref{figure5}
we have seen that the cross section of the scattering with
bound electrons in L shell(2s or 2p states) agrees well with 
the cross section of the scattering with free electron at rest
and the bound state features give small corrections to the scattering
process. It's natural to expect that the bound state features of
electrons in M, N and O shells should give even smaller corrections
compared that of the electrons in L shell. This is easy to understand
because the binding energies of states in these shells are
much smaller than that of the states in L shell, as can be seen
in Table \ref{Tab:Ru-atom}. According to this discussion we
can approximate the cross section of the scattering with electrons
in M, N and O shells as that of the scattering with free electron
at rest.

{\bf Scattering of solar neutrino with bound electron in Ru atom}

\begin{table}
\begin{center}
\begin{tabular}{|c|c|c|}
\hline
Sources & Fluxes ($10^{10}$ cm$^{-2}$ s$^{-1}$)  & Energy(MeV)
\\
\hline
pp & $6.0$ ($\pm 2\%$) & [0, 0.423] 
\\
\hline
pep & $0.014$($\pm 5\%$) & 1.44
\\
\hline
hep & $8.\times 10^{-7}$ & [0, 18.78]
\\
\hline
$^7$Be & $0.47$($\pm 15 \%$) & 0.861
\\
\hline
  $^8$B & $5.8 \times 10^{-4}$  ($\pm 37\%$) & [0, 16.40]
\\
\hline
  $^{13}$N & $0.06$ ($\pm 50 \%$) & [0, 1.199] 
\\
\hline
  $^{15}$O & $0.05$ ($\pm 58 \%$) & [0, 1.732]
\\
\hline
 $^{17}$F & $5.2\times 10^{-4}$($\pm 46\%$) & [0, 1.74]
\\
\hline
\end{tabular}
\end{center}
\caption{\it Calculated solar neutrino flux\cite{bahcall}. Uncertainties
 of solar neutrino fluxes are shown in brackets.}
\label{solarnu}
\end{table}

Solar neutrinos are electron neutrinos produced in fusion reactions
or in decays of radioactive nuclei. As can be seen in Table \ref{solarnu},
solar neutrino fluxes are dominated by the pp neutrino and
$^7$Be neutrino. Fluxes of $^{13}$N neutrinos, $^{15}$O neutrinos
and pep neutrinos are also not small. Solar neutrinos also
have a quite wide spectrum. For $E_\nu \lsim 0.4$ MeV, the
solar neutrino spectrum is dominated by the pp neutrinos.
For $E_\nu \gsim 1.74$ MeV, it is dominated by $^8$B neutrinos.
In between $0.4$ MeV and $1.74$ MeV, major contributions
are from $^7$Be neutrinos, $^{13}$N neutrinos and $^{15}$O neutrinos
~\cite{bahcall}. 

Solar electron neutrinos oscillate into
muon neutrinos and tau neutrinos in propagation from the 
Sun to the Earth~\cite{w,ms}. The probability of electron
neutrinos surviving as electron neutrinos after 
propagation to the Earth depends on the energy of the neutrinos
and on the matter density profile in the Sun.
In ~\cite{Liao0} it was shown that the survival probability of
solar electron neutrinos can be well described using a formula
with an average density associated with the neutrino production
in the Sun. Since eight types of solar neutrinos, 
as shown in Table \ref{solarnu},
have different production distribution in the Sun, the average
survival probabilities are different for different
types of solar neutrinos. For solar neutrino of type k, we
label $P_k$ as the survival probability of the electron neutrinos
after propagation to the Earth. $P_k$ can be found in ~\cite{Liao0}.
The survival probability also depends on the 
mass squared difference, $\Delta m^2_{21}$,
and the vacuum mixing angle $\theta_{12}$.
When computing $P_k$ we use
\bea
\Delta m^2_{21} = 7.62\times 10^{-5} ~\textrm{eV}{^2},
~~sin^2 \theta_{12}=0.32.
\eea

In the previous section we find that
in the scattering with bound electrons
solar neutrinos in a very wide energy range can contribute to
the events with final electron of a kinetic energy $E_k\approx 42$ keV.
In particular, this means that neutrinos with energy less
than $\frac{1}{2}(E_k+\sqrt{E_k^2+2E_k m_e})$, that is $\lsim 126$ keV
for $E_k\approx 42$ keV, can contribute to the events with $E_k\approx 42$ keV.

Using the calculated neutrino fluxes and the energy spectrum
for all eight types of neutrinos given in ~\cite{bahcall},
we can compute the rate of events in the interested energy range, 
i.e. in range $E_k\approx 42$ keV.
The differential event rate of contributions of
neutrinos with energy $E_\nu$ is computed as follows
\bea
\frac{d R}{d E_\nu}
=\frac{d \sigma_{\nu_e}(E_\nu)}{d E_k} ~N_e ~\sum_k F^k_\nu(E_\nu) P_k 
+\frac{d \sigma_{\nu_\mu}(E_\nu)}{d E_k} ~N_e ~\sum_k F^k_\nu(E_\nu) (1-P_k),
\label{rate}
\eea
where $k$ runs over eight types of solar neutrinos,
$F^k_\nu$ the flux distribution of type k solar neutrino, 
$N_e$ the number of electrons in Ru target.
$\sigma_{\nu_e}$ and $\sigma_{\nu_\mu}$ are
the cross sections for scattering of electron neutrino and
muon neutrino separately. $P_k=P_k(E_\nu)$ is the survival probability
of electron neutrino described above.
$1-P_k$ is the probability of solar electron neutrinos oscillating into
muon neutrinos and tau neutrinos. 
Since the scattering of $\nu_\mu$ and 
$\nu_\tau$ with electrons are universal, we use $\sigma_\mu$ in
Eq. (\ref{rate}).
For simplicity, we neglect the
Earth matter effect in our calculation since it would lead
to at most about $4\%$ corrections for large energy solar neutrino~\cite{Liao0}.
For solar neutrinos with energy less about 1 MeV, the Earth
matter effect would be smaller than $1\%$ and can be safely neglected.

The differential cross section used in (\ref{rate})
is an average for the scattering with all electrons
in the neutral Ru atom. It is written as
\bea
\frac{d \sigma_{\nu_e}}{d E_k}=
\frac{1}{Z} \sum_l ~n_l ~\frac{d\sigma_l}{d E_k},
\label{X-sec4}
\eea
where $l$ runs over all states in Table \ref{Tab:Ru-atom},
$n_l$ the number of electrons in state $l$, $Z=44$ for Ru atom. 
For electrons in M, N and O shells we approximate 
$d\sigma_l/dE_k$ as the same as
that of the scattering with free electron at rest, as discussed in
the last section.
Similarly we have an expression for $d\sigma_{\nu_\mu}$. 

The total rate is obtained after integrating Eq. (\ref{rate})
over $E_\nu$. For 10 tons of $^{106}$Ru we obtain
\bea
R=0.21 \times \frac{\Delta E_k}{10 ~\textrm{eV}} ~\textrm{year}^{-1}.
\label{rate0}
\eea
We can see in Eq. (\ref{rate0}) that
if the energy of final electron can be measured
to a resolution of $10$ eV, solar neutrinos
would produce about 0.2 electron events in the
scattering with electrons in 10 tons of $^{106}$Ru target.
In \cite{Liao} it was shown that for $\theta$, the mixing of the 
sterile neutrino DM to electron neutrino, 
of order $\theta^2=10^{-6}$, the $\nu_s$ capture by $^{106}$Ru
can produce tens of events per year. Eq. (\ref{rate0}) says
that background from the scattering with solar neutrinos
allows us to measure $\theta^2$ to a precision of $10^{-7}$
if the energy of the final electron can be measured to
a precision of 10 eV. If the energy of final electron
can only be measured to a resolution of 100 eV, this
background allows us to measure $\theta^2$ to a precision
of $10^{-6}$.

In calculation of the events rate 
we find that the scattering with solar neutrinos of energy $\lsim 126$ keV
gives small contribution to the result given in Eq. (\ref{rate0}).
This is because 1) the cross section in this energy range is
suppressed compared to that in the energy range $E_\nu \gsim 126$ keV,
as can be seen in Figs. \ref{figure4} and \ref{figure5};
2) the solar neutrinos in this energy range only account
for about $7\%$ of the total solar neutrino flux.
We note that electron after passing out of the target
can lose energy and create broadening of spectrum. But this
does not change our result in Eq. (\ref{rate0}) because
the event rate R varies very slowly with $E_k$. A spectrum
broadening of $10-100$ eV level just gives rise to a mild
re-distribution of events in the original spectrum and it would 
not change the estimate of a continuous spectrum in a range
as wide as a few to around ten keV.

{\bf Conclusion:}

In summary we have studied in detail the scattering of
solar neutrinos with bound electrons in Ru atom.
This study is helpful to clarify the background events
caused by solar neutrinos in the search of keV scale
sterile neutrino DM using $^{106}$Ru target.

We concentrate on the scattering of solar neutrinos
with electrons in the 1s, 2s and 2p states in Ru atom.
We find that for small $E_\nu$ the scattering of neutrinos
with free electron at rest and the scattering with
bound electron can be quite different. For large $E_\nu$,
the difference tends to be small. For $E_\nu > m_e$,
the difference tends to be negligible.
For events of final electrons with a fixed kinetic energy,
say $E_k\approx 42$ keV, we find that the scattering of
neutrino with free electron at rest starts to contribute
when $E_\nu \gsim 126$ keV. On the other hand, the
scattering of neutrino with bound electron
starts to contribute from values of $E_\nu$ much smaller
than $126$ keV. This means that
low energy part of the solar neutrino spectrum can contribute
to the scattering. This part of solar neutrino spectrum
would be neglected when using the cross section of the
scattering of neutrino with free electrons at rest.
Fortunately, solar neutrinos with energy $E_\nu \lsim 126$ keV
only account for about $7\%$ of the total neutrino flux and 
the scattering with
bound electrons does not give a large difference
compared to that computed using the scattering with free electron
at rest.

We estimate the event rate of electrons produced by the scattering
of solar neutrinos with electrons in Ru atom. We find that events
of final electrons having $E_k\approx 42$ keV are $0.2$ 
per year if
the energy of the final electron can be measured to a precision
of $10$ eV. This allows to search for the $\nu_s$ DM
with a mixing of $\nu_s$ and $\nu_e$ at order $\theta^2=10^{-7}$.
If the energy of the final electron can be measured to a precision
of $100$ eV, it allows to search for the $\nu_s$ DM
with a mixing of $\nu_s$ and $\nu_e$ at order $\theta^2=10^{-6}$.
For 10 kg $^3$T as used in \cite{Liao} for the search of $\nu_s$ DM,
the rate of this type of background events is smaller by about three
orders of magnitude. It does not create problem
for the search of $\nu_s$ DM using $^3$T target.
We find that for larger energy resolution, the event rate
of the background is larger. To avoid the pollution of this type of electron
events in the search of sterile neutrino DM, we should have 
good energy resolution to suppress this type of background events. 

\acknowledgments
This work is supported by National Science Foundation of
 China(NSFC), grant No.11135009, No. 11375065 and Shanghai Key Laboratory
 of Particle Physics and Cosmology, grant No. 11DZ2230700.


\begin{thebibliography}{99}

\bibitem{structure}
J. Sommer-Larsen, A. Dolgov,
Astrophys. J. {\bf 551}, 608(2001);

P. Colin, V. Avila-Reese, O. Valenzuela,
Astrophys. J. {\bf 542}, 622(2000).

\bibitem{Cusp}
C. Destri, H. J. de Vega and N. G. Sanchez, New Astron. {\bf 22}, 39(2013);
Astrop. Phys. {\bf 46}, 14 (2013).

\bibitem{Liao}
W. Liao, Phys. Rev. D{\bf 82}, 073001(2010).

\bibitem{Liao2}
W. Liao, 
in Highlights and Conclusions of the Chalonge CIAS Meudon Workshop 2012,
eds. P.L. Biermann, H. J. de Vega, N.G. Sanchez,
arXiv: 1305.7452.

\bibitem{ABS} 
 T. Asaka, S. Blanchet and M. Shaposhnikov,
 Phys. Lett. B{\bf 631}, 151(2005).

\bibitem{review2}
For recent reviews, see A. Merle, Int. J. Mod. Phys. D{\bf 22}, 1330020(2013);
M. Drewes, Int. J. Mod. Phys. E{\bf 22}, 1330019(2013).

\bibitem{DW} 
 S. Dodelson and L. W. Windrow,
 Phys. Rev. Lett. {\bf 72}, 17(1994).

\bibitem{others} A. D. Dolgov and S. H. Hansen, Astropart. Phys. {\bf 16}, 339 (2002);
K. Abazajian, G.M. Fuller, and M. Patel, Phys. Rev. {\bf D64}, 023501 (2001);
K. Abazajian, G.M. Fuller, and W.H. Tucker, Astrophys. J. {\bf 562}, 593 (2001).

\bibitem{blrv} 
 A. Boyarsky, et.al., Phys. Rev. Lett. {\bf 102}, 201304(2009).

 \bibitem{ST} 
 M. Shaposhnikov and I. Tkachev, Phys. Lett. {\bf B639}, 414 (2006).

\bibitem{PK}
 K. Petraki and A. Kusenko, Phys. Rev. {\bf D77}, 065014 (2008).

  \bibitem{bhl}
 F. Bezrukov, H. Hettmansperger, M. Lindner, Phys. Rev. {\bf
 D81}, 085032(2010).

  \bibitem{bnrst} 
 A. Boyarsky, et.al., Phys. Rev. Lett. {\bf 97}, 261302(2006).

 \bibitem{bnr} 
 A. Boyarsky, J. Nevalainen and O. Ruchayskiy, Astron. Astrophys. {\bf 471}, 51(2007).

\bibitem{birs}
 A. Boyarsky, D. Iakubovskyi, O. Ruchayskiy and V. Savchenko,
 MNRAS {\bf 387}, 1361(2008).

\bibitem{blrv2}
  A. Boyarsky, J. Lesgourgues, O. Ruchayskiy, M. Viel,
 JCAP {\bf 0905}, 012(2009).

\bibitem{bri}  
 A. Boyarsky, O. Ruchayskiy, D. Iakubovskyi, JCAP {\bf 0903}, 005(2009).

\bibitem{sh}
 M. Shaposhnikov, Nucl. Phys. B{\bf 763}, 49(2007).

\bibitem{LMN}
M. Lindner, A. Merle, V. Niro,
JCAP {\bf 1101} (2011) 034.

\bibitem{GT}
C.-Q. Geng, R. Takahashi, Phys. Lett. B{\bf 710}, 324(2012).

\bibitem{CDS}
L. Canetti, M. Drewes, M. Shaposhnikov,
Phys. Rev. Lett. {\bf 110}, 061801(2013).

\bibitem{RT}
D. J Robinson, Y. Tsai, JHEP {\bf 1208} (2012) 161.

\bibitem{indirect}
A. Boyarsky, D. Iakubovskyi, O. Ruchayskiy,
Phys. Dark Univ. {\bf 1}, 136(2012).

\bibitem{LX}
Y. F. Li and Z. Z. Xing, JCAP {\bf 1108}(2011)006.

\bibitem{LX1}
Y. F. Li and Z. Z. Xing, Phys. Lett. B{\bf 695}, 205(2011).

\bibitem{SV}
H. de Vega, et.al, 
Nucl. Phys. B{\bf 866}, 177 (2013). 

 \bibitem{BS} 
 F. Bezrukov and M. Shaposhnikov, Phys. Rev. {\bf D75}, 053005(2007).

  \bibitem{ak} 
 S. Ando and A. Kusenko,  Phys. Rev. {\bf D81}, 113006(2010).

\bibitem{CRC} David R. Lide, ed., CRC Handbook of Chemistry and Physics, 
CRC Press, Boca Raton, FL, 2005

\bibitem{X-ray-evaluation}
J. A. Bearden and A. F. Burr, Rev. Mod. Phys. {\bf 39}, 125(1967).

\bibitem{RPP}
 Review of Particle Physics, Phys. Rev. D{\bf 86}, 010001(2012).

\bibitem{PWIA}
M. A. Coplan et al., Rev. Mod. Phys. {\bf 66}, 985(1994).

 \bibitem{bahcall} Neutrino Astrophysics, J. N. Bahcall,
 Cambridge University Press 1989; see also online data at
 website: http://www.sns.ias.edu/~jnb/SNdata/sndata.html

\bibitem{w}
L. Wolfenstein, Phys. Rev. {\bf D 17}, 2369 (1978); L. Wolfenstein, in
"Neutrino-78", Purdue Univ. C3 - C6, (1978).

\bibitem{ms}
S. P. Mikheyev and A. Yu. Smirnov, Yad. Fiz. {\bf 42}, 1441 (1985) [
Sov. J. Nucl. Phys. {\bf 42}, 913 (1985)]; Nuovo Cim. {\bf C9}, 17
(1986); S. P. Mikheyev and A. Yu. Smirnov, ZHETF, {\bf 91}, (1986),
[Sov. Phys. JETP, {\bf 64}, 4 (1986)] (reprinted in "Solar neutrinos:
the first thirty years", Eds. J.N.Bahcall {\it et. al.}).

\bibitem{Liao0}
P. C. de Holanda, W. Liao, A. Yu. Smirnov,
Nucl. Phys. B {\bf 702}, 307(2004).

\end{thebibliography}
\end{document}